%% file: main.tex
\documentclass[a4paper]{jpconf}

\usepackage{graphicx}
\usepackage{amsfonts}
\usepackage{placeins}
\usepackage{subcaption}
\usepackage{color, xcolor, colortbl}
\usepackage{graphicx,epstopdf}
\usepackage{geometry}
\usepackage{amsmath,amssymb,amsthm}
\usepackage{algorithm}
\usepackage{bm}
\usepackage{appendix}
\usepackage{multirow}
\usepackage{braket}
\usepackage[english]{babel}
\usepackage[capitalize]{cleveref}
\usepackage[square,numbers]{natbib}
\usepackage{adjustbox}
\usepackage{xspace}

\newcommand{\bvec}[1]{\mathbf{#1}}

\newcommand{\vr}{\bvec{r}}

\newcommand{\vG}{\bvec{G}}

\newcommand{\vR}{\bvec{R}}

\newcommand{\I}{\mathrm{i}}

\newcommand{\mc}[1]{\mathcal{#1}}

\newcommand{\abs}[1]{\left\lvert#1\right\rvert}

\newcommand{\ud}{\,\mathrm{d}}
\renewcommand{\Or}{\mathcal{O}}

\newcommand{\ZZ}{\mathbb{Z}}

\theoremstyle{plain}

\theoremstyle{plain}

\theoremstyle{plain}
\newtheorem*{lem*}{\protect\lemmaname}
\theoremstyle{plain}

\theoremstyle{plain}

\providecommand{\definitionname}{Definition}
\providecommand{\assumptionname}{Assumption}
\providecommand{\corollaryname}{Corollary}
\providecommand{\lemmaname}{Lemma}
\providecommand{\propositionname}{Proposition}
\providecommand{\remarkname}{Remark}
\providecommand{\theoremname}{Theorem}

\usepackage{bbm}
\usepackage{algpseudocode}
\usepackage{algorithm}
\usepackage{algorithmicx}

\newcommand{\GG}{\mathbb{G}}
\newcommand{\XX}{\mathbb{X}}

\newcommand{\nn}{\mathrm{nn}}

\newcommand{\per}{\mathrm{per}}

\newcommand{\dOmega}{\left\lvert \delta \Omega \right\rvert}

\usepackage{xr}
\makeatletter
\newcommand*{\addFileDependency}[1]{% argument=file name and extension
  \typeout{(#1)}
  \@addtofilelist{#1}
  \IfFileExists{#1}{}{\typeout{No file #1.}}
}
\makeatother

\newcommand*{\myexternaldocument}[1]{%
    \externaldocument{#1}%
    \addFileDependency{#1.tex}%
    \addFileDependency{#1.aux}%
}

\myexternaldocument{SupplementaryMaterial}

\begin{document}

\title{Discontinuous Galerkin method with Voronoi partitioning for Quantum Simulation of Chemistry}

\author{Fabian M. Faulstich}
\address{Department of Mathematics, University of California, Berkeley,  CA 94720, USA}
\author{Xiaojie Wu}
\address{Department of Mathematics, University of California, Berkeley,  CA 94720, USA}
\author{Lin Lin}
\address{Department of Mathematics, University of California, Berkeley,  CA 94720, USA}
\address{Computational Research Division, Lawrence Berkeley National Laboratory, Berkeley, CA 94720, USA}
\ead{linlin@math.berkeley.edu}

\begin{abstract}
Molecular orbitals based on the linear combination of Gaussian type orbitals are arguably the most employed discretization in quantum chemistry simulations---both on quantum and classical devices. To circumvent a potentially dense two-body interaction tensor and obtain lower asymptotic costs for quantum simulations of chemistry, the discontinuous Galerkin (DG) procedure using a rectangular partitioning strategy was recently piloted [McClean et al, New J. Phys. 22, 093015, 2020]. The DG approach interpolates in a controllable way between a compact description of the two-body interaction tensor through molecular orbitals and a diagonal characterization through primitive basis sets, such as a planewave dual basis set. The DG procedure gives rise to a block-diagonal representation of the two-body interaction with reduced number of two-electron repulsion integrals, which in turn reduces the cost of quantum simulations. In the present work we extend this approach to be applicable to molecular and crystalline systems of arbitrary geometry. We take advantage of the flexibility of the planewave dual basis set, and combine the discontinuous Galerkin procedure with a general partitioning strategy based on the Voronoi decomposition. We numerically investigate the performance, at the mean-field and correlated levels, with quasi-1D, 2D and 3D partitions using hydrogen chains, H$_4$, CH$_4$ as examples, respectively. We also apply the method to graphene as a prototypical example of  crystalline systems.  
\end{abstract}

\section{Introduction}

Over the past decades, the ability to obtain solutions to the electronic structure problem for systems containing hundreds of atoms, or even more, has revolutionized chemistry allowing to address important problems through computational efforts. 
Today, quantum computation is a promising candidate to push the frontiers of \textit{ab initio} quantum chemistry even further. 

Arguably the most pronounced challenge of using quantum devices today and in the near future is the short coherence time of qubits that requires low circuit-depth algorithms, i.e. quantum algorithms that are executable in as little time as possible.
Circuit depth refers to the number of layers of simultaneous gates used to compose a circuit. It is used to measure time complexity and is limited by the coherence time. 
%
%For variational algorithms the circuit depth is a crucial metric for assessing viability~\cite{}.
%
%
% This is problematic since many algorithms have computational costs that scale explicitly with the number of terms~\cite{babbush2018low}.
% %
% To that end, we enhance the development of the {\it discontinuous Galerkin} formalism for simulations of quantum chemistry on quantum computers~\cite{mcclean2020discontinuous}---a new approach to the basis-set design for quantum simulations of chemistry.
%
While the design of basis sets for classical simulations of quantum chemistry is mainly driven by finding the most compact description of the Hamiltonian~\cite{helgaker2014molecular}, the ideal basis set for quantum simulations of quantum chemistry exhibits a certain level of sparsity~\cite{babbush2018low}.   
Among various basis sets available, most classical implementations of quantum chemistry methods use Gaussian type orbitals (GTOs) in a molecular orbital formulation.
These basis sets offer compact representations of correlated problems and also present an energy ordering of orbitals that facilitates further reduction of the space through active-space methods. 
A side effect of this  compact basis set is that molecular orbitals are delocalized in the real space, and this leads to an almost dense description of the Hamiltonian. In particular, the number of entries in the two-electron repulsion interaction (ERI) tensor scales nominally as  $\mathcal{O}(N^4)$, where $N$ is the number of basis functions.  
In the context of quantum simulation, the large number of entries in the ERI tensor can significantly increase the circuit depth and hence the computational cost.
Moreover, GTOs tend to be nearly linearly dependent, especially when diffuse basis functions are considered. This can commonly result in an ill-conditioned overlap matrix and consequently convergence issues already at the level of the Hartree--Fock theory.  

Alternatively, primitive basis sets (also referred to as numerical basis sets~\cite{lehtola2019review}) such as the planewave dual basis set offer sparse representations of the ERI tensor due to a grid-based representation, and the ERI tensor can be mapped to a diagonal matrix. This diagonal representation effectively reduces the asymptotic scaling of quantum algorithms~\cite{babbush2018low,KivlichanMcCleanWiebeEtAl2018}. However, the grid based representation also implies that in order to perform accurate calculations  (with an error of $\sim 1\,{\rm mH}$), the number of basis functions can be several orders of magnitude larger than that of GTOs or molecular orbitals, particularly in the context of all-electron calculations. 

In a recent work~\cite{mcclean2020discontinuous}, we have introduced a {\it Discontinuous Galerkin} (DG) inspired basis set to interpolate between the compact but dense description of the Hamiltonian through GTOs / Gaussian-based molecular orbitals, and the sparse description by means of a primitive basis set.
In the DG formalism, the ERI tensor can be mapped to a block-diagonal matrix, and hence quantum simulation performed using the DG basis enjoys the same reduced asymptotic scaling as that using the primitive basis set. Meanwhile, the span of the DG basis set is larger than that of the Gaussian basis functions, and can capture therefore correlation effects with a relatively small overhead. In particular, the number of DG basis functions can be orders of magnitude lower than that of the primitive basis functions.

In~\cite{mcclean2020discontinuous}, the computational domain is always partitioned into a set of rectangular subdomains, which is ideally suited for describing quasi-1D systems such as hydrogen chains. The rectangular partitioning strategy is also consistent with the use of the DG basis functions for density functional theory (DFT) based calculations~\cite{lin2012adaptive,HuLinYang2015a}, and will be referred to as the DG-Rectangular method (or {DG-R} for short). We find that when atoms are placed near the boundary of a subdomain, the error can increase compared to when the atoms appear in the center of a subdomain. This error eventually diminishes as the basis set approaches the complete basis set limit. However, this can increase the overhead of quantum simulation with a relative small basis set. One drawback of the {DG-R} method is that even for simple molecular configurations, it is not always possible to ensure that all atoms are located at the interior of the subdomain. 

The main contribution of this paper is to present a simple solution to the problem above. Our method is based on the observation that the planewave dual basis set can be identified with a discrete set of grid points that can be partitioned into \textit{arbitrary} subsets, without increasing the complexity of the algorithm or the computational cost. 
This allows us to partition the computational domain into subdomains of polygons of arbitrary shapes, so that all atoms are located at the interior of a subdomain.
This is achieved using a partitioning strategy based on the Voronoi decomposition. %
We refer to this procedure as the discontinuous Galerkin formalism with Voronoi partitioning (DG-Voronoi or {DG-V} for short).
For quasi one-dimensional systems with a linear geometry, the {DG-R} and {DG-V} partitioning strategies coincide with each other. However, the {DG-V} approach is much more flexible and naturally generalizes to higher-dimensional molecular geometries.

We start by introducing the discretization of the electronic-structure problem in a planewave dual basis, and define precisely what is meant by a diagonal  representation of the ERI tensor and strictly localized functions. 
The discontinuous Galerkin approach for an arbitrary partitioning is introduced as a general framework for maintaining these properties by using the planewave dual basis set as a primitive building block. 
We then introduce the Voronoi partitioning strategy, which will be used in practical application to obtain a partitioning of the supercell for molecular and crystalline systems in the discontinuous Galerkin procedure.  We implement the {DG-V} method (together with a new implementation of the {DG-R} method) using the PySCF software package~\cite{sun2018pyscf}. The performance of the {DG-V} method is demonstrated using molecular systems including H$_4$, hydrogen chain, CH$_4$, as well as a periodic graphene system.

% %Rewrite:
% We show that the {DG-V}oronoi approach both maintains accuracy in correlated calculations, and demonstrates a crossover to a lower number of non-zero two electron interaction integrals at modest system sizes between 12 and 14 atoms.
% %
% %We finish with an outlook on how this approach will influence quantum and classical approaches to correlated electronic structure alike
%  

\section{Theory}
\label{sec:Method}

The {DG-R} formalism described in Ref.~\cite{mcclean2020discontinuous} interpolates between a primitive basis set and an active space basis set.
As primitive basis set we here investigate the use of planewave dual functions. 
The reason for this is twofold:
First, the planewave dual functions are grid based nascent delta functions, which trivializes the characterization of the basis projection matrix. 
Second, the {\it periodic boundary condition} module allows an implementation relying solely on PySCF---a program suit for electronic structure calculations in Python~\cite{sun2018pyscf}.%---as a third-party electronic-structure software.   

\subsection{Planewave dual basis set and the active space basis set}

The many-body Hamiltonian for an isolated system with $N$ electrons and $M$ nuclei is given by 
\begin{equation} 
  \hat{H} = -\sum_{i=1}^N \frac12 \nabla_{\vr_i}^2 - \sum_{i=1}^N \sum_{I=1}^M
\frac{Z_I}{|\vr_i-\vR_I|} + \sum_{1\le i < j \le N}
\frac{1}{|\vr_i - \vr_j|} + \sum_{1\le I < J \le M}\frac{Z_IZ_J}{|\vR_I-\vR_J|},
\label{eqn:manybody_allelectron_freespace}
\end{equation}
where we have assumed atomic units, the Born--Oppenheimer approximation
such that the positions of nuclei $\vR_I$ are constants, $\mathbf{r}_i$ represent the positions of electrons, and $Z_I$ is the charge of the $I$-th nuclei. We assume the system is charge neutral, i.e. $\sum_{I=1}^M Z_I=N$.

The practical simulation of electronic structure problems often uses the supercell model. For simplicity, the computational domain $\Omega$ is defined to be an orthorhombic cell $[0,L_1]\times[0,L_2]\times[0,L_3]$, with periodic boundary conditions. Its volume is denoted by $\abs{\Omega}=L_1L_2L_3$. The supercell defines a Bravais lattice
\begin{equation}
\mathbb{L} = \left\{ (L_1 i_1, L_2 i_2, L_3 i_3),
\quad (i_1,i_2,i_3)\in \ZZ^3 \right\}.
\end{equation}

The supercell setup is necessary for simulating solid state systems. Even for molecular systems, the supercell model is also widely used, particularly in the context of density functional theory calculations \cite{MakovPayne1995,KresseFurthmuller1996}. In the second quantized formulation, the quantum many-body Hamiltonian in the supercell model in a planewave dual basis set takes the form

\begin{equation}
\hat{H}=\sum_{\mu,\nu=1}^{N_p} h^{(p)}_{\mu\nu} \hat{b}_{\mu}^{\dagger}
  \hat{b}_{\nu} + \frac12 \sum_{\mu,\nu=1}^{N_{p}} v^{(p)}_{\mu\nu}   
  \hat{n}_{\mu} \hat{n}_{\nu} + E_{\nn}(\{\vR_I\}),
\label{eqn:manybody_pwdual}
\end{equation}
where 
\begin{equation}
h^{(p)}_{\mu\nu}=\frac12 \int_{\Omega} \nabla\overline{\chi}_{\mu}(\vr) \cdot \nabla \chi_{\nu}(\vr) \ud \vr - \sum_{I=1}^M
Z_I v^{\per}(\vr_{\mu}-\vR_I) \delta_{\mu\nu},
\end{equation}
and
\begin{equation} 
v^{(p)}_{\mu\nu}=v^{\per}(\vr_{\mu}-\vr_{\nu}).
\end{equation}
The nuclei-nuclei interaction takes the form
\begin{equation} 
E_{\nn}(\{\vR_I\})=\frac12 \sum_{I=1}^M \sum_{J\ne I} Z_I Z_J v^{\per}(\vR_{I}-\vR_J).
\end{equation}
Here $\{\chi_{\mu}(\vr)\}_{\mu=1}^{N_p}$ denotes the planewave dual basis set, $\hat{b}_{\mu}^{\dagger},\hat b_{\mu}$ are the associated creation and annihilation operator, and $\hat{n}_{\mu}=\hat{b}_{\mu}^{\dagger}\hat{b}_{\mu}$ is the number operator. Note that the Coulomb kernel should be replaced by a properly shifted Ewald potential denoted by $v^{\per}(\vr-\vr')$. In the following discussions, all basis sets will be represented as linear combination of the planewave dual basis set. Hence the superscript $(p)$ indicates that the planewave dual basis set is a ``primitive'' basis set. We refer readers to \ref{app:planewavedual} for more details. Here the number of primitive basis functions $N_p$ is equal to the number of grid points $N_g$ as in \cref{eqn:number_grid}.

From \cref{eqn:ERI_diagonal}, the electron repulsion integral (ERI) tensor in the planewave dual basis set can be viewed as a diagonal matrix. Therefore the Hamiltonian of the form \cref{eqn:manybody_pwdual} is called a ``diagonal Hamiltonian''. For diagonal Hamiltonians, the number of nonzero terms scales quadratically with respect to the number of basis functions as $\Or(N_p^2)$.
For quantum simulation, the planewave dual basis set allows the efficient implementation using ``fermionic fast Fourier transform'' (FFFT), and ``fermionic swap networks'' to reduce the asymptotic complexity. We refer readers to \cite{babbush2018low,KivlichanMcCleanWiebeEtAl2018} for more details.

The planewave basis set can be systematically improved by increasing the number of grid points $N_p$. The main disadvantage is that $N_p$ can be very large. Even in the pseudopotential formulation, the ratio $N_p/N$ can often be $10^3\sim 10^4$. Therefore the direct usage of the planewave dual basis set for correlated electronic structure calculations, as well as practical simulation on quantum computers can be  prohibitively expensive. 
As a result, quantum chemistry calculations commonly utilize physically motivated basis sets that result in a more compact description (e.g. the Pople basis sets such as X-YZg split-valence basis sets~\cite{ditchfield1971self} or the correlation consistent basis sets~\cite{dunning1989gaussian} in the LCAO approach~\cite{helgaker2014molecular}). 
Let $\{\varphi_{p}(\vr)\}_{p=1}^{N_{a}}$ be a set of orthonormal single-particle functions; we refer to them as ``active space'' orbitals (for instance, canonical Hartree--Fock orbitals, or natural orbitals), which can be expanded using a primitive basis set as
\begin{equation}
  \varphi_{p}(\vr) = \dOmega^{\frac12} \sum_{\mu} \chi_{\mu}(\vr) \Phi_{\mu p},
  \label{eq:GramMatrix}
\end{equation}
where $\Phi\in \mathbb{C}^{N_{p}\times N_{a}}$ is a coefficient matrix with orthogonal columns, with entries given by the nodal representation as $\Phi_{\mu p}=\varphi_p(\vr_\mu)$, and   $\dOmega:=|\Omega|/N_g$ the size of the volume element. 
  For flexibility and accuracy, we will allow quite general definitions of the active space basis set in this work.  
It will pertain to traditional Hartree-Fock canonical orbitals as well as Gaussian basis sets such as the Dunning cc-pVDZ basis set \cite{dunning1989gaussian}.
%, which allows us to use Gaussians with strictly block diagonal properties \LL{ what does Gaussians with strictly block diagonal properties mean?} by going through the primitive basis set using point sampling through the approximate delta function properties of the primitive basis set. 
The main advantage of using an active space basis is that we may have $N_a\ll N_p$, and the ration $N_a/N$ is usually $10\sim 100$.

Taking as granted the construction of the matrix $\Phi$, we define a rotated set of creation and annihilation operators in the active space as 
\begin{equation}
  \hat{a}^{\dagger}_{p} = \sum_{\mu=1}^{N_p} \hat b^{\dagger}_{\mu} \Phi_{\mu p}, \hspace{2.5cm} \hat{a}_{p} = \sum_{\mu=1}^{N_p} \hat  b_{\mu}
  \overline{\Phi}_{\mu p},
  \label{eqn:smallbasis_op}
\end{equation}
where $\overline{\Phi}_{\mu p}$ denotes the complex conjugate and we may project the Hamiltonian as
\begin{equation}
  \hat{H}^{(a)} = \sum_{p,q=1}^{N_a} h^{(a)}_{pq} \hat{a}_p^\dagger \hat{a}_q + \frac{1}{2} \sum_{p,q,r,s=1}^{N_a} v^{(a)}_{pqrs} \hat{a}_p^\dagger \hat{a}_q^\dagger \hat{a}_r \hat{a}_s,
\end{equation}
which we refer to as the active space Hamiltonian. 

Generally we see that the  primitive basis sets have favorable scaling in number of terms in the Hamiltonian ($\mc{O}(N_p^2)$ vs $\mc{O}(N_a^4)$), which often corresponds to better scaling algorithms.  
While $N_p$ and $N_a$ have the same asymptotic scaling, for modest sized calculations it is often observed that $N_p \gg N_a$ in order to achieve comparable accuracy.  
Here we will seek a way to split the difference between these two regimes by forming a more compact basis that partially retains the diagonal properties, i.e. the resulting Hamiltonian is block-diagonal.

\subsection{Discontinuous Galerkin basis set}
\label{sec:DG-general}

The core idea of Ref.~\cite{mcclean2020discontinuous} is to construct the block-diagonal basis by fitting spatially connected blocks of the primitive basis set to the active space basis set, while preserving the properties of the primitive basis set.
The DG approach can therefore be seen as an interpolation between the primitive basis set and the active space basis set.  

The DG approach systematically compress the active space basis set $\{\varphi_{p}(\vr)\}_{p=1}^{N_{a}}$ into a set of orthonormal basis functions partitioned into spatial elements---the DG elements.
We enforce that the basis functions associated with different DG elements have mutually disjoint supports.
Assume that the index set $\Omega=\{1,\ldots,N_{p}\}$ can be partitioned into $N_b$ index sets (blocks), i.e.
\begin{equation}
  \mathcal{K} = \{\kappa_1, \kappa_2, \cdots, \kappa_{N_b} \}.
\end{equation}
The basis-projection matrix $\Phi$ can then be partitioned into $N_b$ blocks $\Phi_{\kappa} := [\Phi_{\mu p}]_{\mu \in \kappa}$ for $\kappa\in\mathcal{K}$. 
The low rank approximation of $\Phi_{\kappa}$, i.e.
\begin{equation}
\label{eq:low-rank-phi}
\Phi_{\kappa} \approx U_{\kappa} S_{\kappa} V_{\kappa}^{\dagger},
\end{equation}
where $U_{\kappa}$ is a matrix with orthonormal columns corresponding to
the leading $n_{\kappa}$ singular values up to some truncation tolerance
$\tau$, yields a compressed basis
\begin{equation}
\phi_{\kappa,j}(\vr) = \dOmega^{\frac12}\sum_{\mu \in \kappa} \chi_{\mu}(\vr)(U_{\kappa})_{\mu j}.
\end{equation}
Again, the coefficient matrix can be obtained from the nodal values as $(U_{\kappa})_{\mu j}=\phi_{\kappa,j}(\vr_{\mu})$. 

Note that the matrix $\Phi$ representing the nodal values of the active space basis set does not need to be orthogonal.  This is in particular the case, when the active space basis is chosen to be a Gaussian type atomic orbital basis. Yet, the compressed DG basis $\phi_{\kappa,j}$ is \emph{always} orthogonal as $U_\kappa$ is orthonormal for all $\kappa$ and as DG basis functions of different DG elements have disjoint (discrete) support sets.

The basis set is adaptively compressed with respect to the given set of basis functions, and are locally supported (in a discrete sense) only on a single index set $\kappa$.
We refer to this basis set $\{\phi_{\kappa,j}\}$ as the DG basis set. 
Note that each DG basis function $\phi_{\kappa,j}$ is a linear combination of primitive basis functions which are themselves continuous, so $\phi_{\kappa,j}$ is also \textit{technically} continuous in real space.
In fact, $\phi_{\kappa,j}$ might not be locally supported in  real space if each primitive basis function $\chi_{p}$ is delocalized. 
When the primitive basis functions are localized, $\phi_{\kappa,j}$ can be very close to a discontinuous function (see Figure~\ref{fig:DGV-funcations}). 
When computing the projected Hamiltonian, we do not need to evaluate the surface terms in the standard DG formalism~\cite{Arnold1982,lin2012adaptive}.  
Furthermore, in the absence of truncation, we clearly have $\text{span} \{\varphi_{p}\}\subseteq \text{span} \{\phi_{\kappa,j}\}$.

For the block diagonal matrix
\begin{equation}
U 
%= \text{diag}[U_{1},\ldots,U_{N_b}]
=
\begin{pmatrix}
U_1 & 0 & \cdots & 0\\
0 & U_2 & \ddots & \vdots \\
\vdots & \ddots & \ddots & 0\\
0 & \cdots & 0 & U_{N_b}\\
\end{pmatrix}
,
\label{}
\end{equation}
the total number of basis functions is thus given by $N_{d}:=\sum_{\kappa\in\mathcal{K}} n_{\kappa}$.
We remark that the number of basis functions $n_{\kappa}$ can in principle be different across different elements. 
We can furthermore define a new set of creation and annihilation operators that correspond to the DG basis functions, i.e.
\begin{equation}
\hat{c}^{\dagger}_{\kappa,j} 
= \sum_{\mu} \hat{b}^{\dagger}_{\mu} (U_{\kappa})_{\mu j}, 
\hspace{2.5cm} 
\hat{c}_{\kappa,j} = \sum_{\mu} \hat{b}_{\mu}(\overline{U}_{\kappa})_{\mu j},
\label{eqn:dalb_op}
\end{equation}
with $\kappa=1,\ldots,N_{b}$ and
$j=1,\ldots,n_{\kappa}$.
Note that unlike Eq.~\eqref{eqn:smallbasis_op}, the basis set rotation in
Eq.~\eqref{eqn:dalb_op} is restricted to the individual DG-elements. 
We readily obtain the projected Hamiltonian in the DG basis as
\begin{equation}
\hat{H}^{(d)} 
= \sum_{\kappa,\kappa';j,j'} h^{(d)}_{\kappa,\kappa';j,j'}
  \hat{c}^{\dagger}_{\kappa,j} \hat{c}_{\kappa',j'} +
  \frac12 \sum_{\kappa,\kappa';i,i',j,j'}
  v^{(d)}_{\kappa,\kappa';i,i',j,j'}
  \hat{c}^{\dagger}_{\kappa,i} \hat{c}^{\dagger}_{\kappa',i'}
  \hat{c}_{\kappa',j'} \hat{c}_{\kappa,j}.
  \label{eqn:H_DG}
\end{equation}
The matrix elements are
\begin{subequations}
\begin{align}
h_{\kappa,\kappa' ; j,j'}^{(d)} &= \sum_{\mu \nu} (\overline{U}_{\kappa})_{\mu j}
h^{(p)}_{\mu \nu} (U_{\kappa'})_{\nu j'},\\
v_{\kappa,\kappa' ; i,i',j,j'}^{(d)} &= 
\sum_{\mu \nu } (\overline{U}_{\kappa})_{\mu i}
(\overline{U}_{\kappa'})_{\nu i'} v^{(p)}_{\mu \nu}
(U_{\kappa})_{\mu j} (U_{\kappa'})_{\nu j'}.
\end{align}
\end{subequations}

Generally the one-body matrix $h^{(d)}$ can be a full dense matrix, but
the two-body interaction tensor $v^{(d)}$ always takes a ``block diagonal'' form in the following sense (it has a specific sparsity pattern). 
In principle, the two-body interaction tensor in the DG basis set should take the form
\begin{equation} 
\frac12 \sum_{\kappa,\kappa',\lambda,\lambda';i,i',j,j'}
  v_{\kappa,i;\kappa',i';\lambda,j;\lambda',j'}
  \hat{c}^{\dagger}_{\kappa,i} \hat{c}^{\dagger}_{\kappa',i'}
  \hat{c}_{\lambda',j'} \hat{c}_{\lambda,j}.
\end{equation}
Compared to Eq.~\eqref{eqn:H_DG}, we find that
\begin{equation}
v_{\kappa,i;\kappa',i';\lambda,j;\lambda',j'} =
v^{(d)}_{\kappa,\kappa';i,i',j,j'} \delta_{\kappa\lambda}
\delta_{\kappa'\lambda'}.
\end{equation}
In other words, $v$ can be viewed as a block diagonal matrix with
respect to the grouped indices $(\kappa\kappa',\lambda\lambda')$.

We remark that the convergence of the DG basis set is independent of the choice of the primitive basis set so long as the primitive basis has sufficient degrees of freedom to form a good approximation to the active space functions of interest. 
At the end of this adaptive procedure, we expect the number of elements in the Hamiltonian to scale as $O(N_b^2 n_{\kappa}^4)$, where we have assumed without loss generality that $n_{\kappa}$ is a constant and that
$N_{d}=N_{b} n_{\kappa}$.  

\subsection{Discontinuous Galerkin basis with Voronoi partitioning}
\label{sec:voronoi}

% In order to apply the above described DG formalism, we require a nodal basis set that is (to a certain extent) localized in real space, and that allows a controllable approximation to the considered active-space basis. 
% %
% An appropriate choice is given by the planewave dual basis set.
% %
% Assuming the grid spacing is fine enough, the nodal nature of the planewave dual basis supports arbitrary partitioning of the considered supercell. 
% %and therewith provides sufficient flexibility to optimally partition the \FF{considered supercell (?)} molecular system.
% %
% In particular, the planewave dual basis set supports the general partitioning via the Voronoi decomposition, which allows us to extend the {DG-R} approach presented in Ref.~\cite{mcclean2020discontinuous}.
% %and therewith to systems of arbitrary geometry. 

% The rectangular partitioning of the underlying domain, which was well-motivated for quasi one-dimensional systems may---to a certain extent---be applicable for a small set of higher dimensional geometries, however, it lacks the general flexibility to describe molecular systems of highly intricate geometries.

\begin{figure}[h!]
    \centering
    \includegraphics[width = \textwidth]{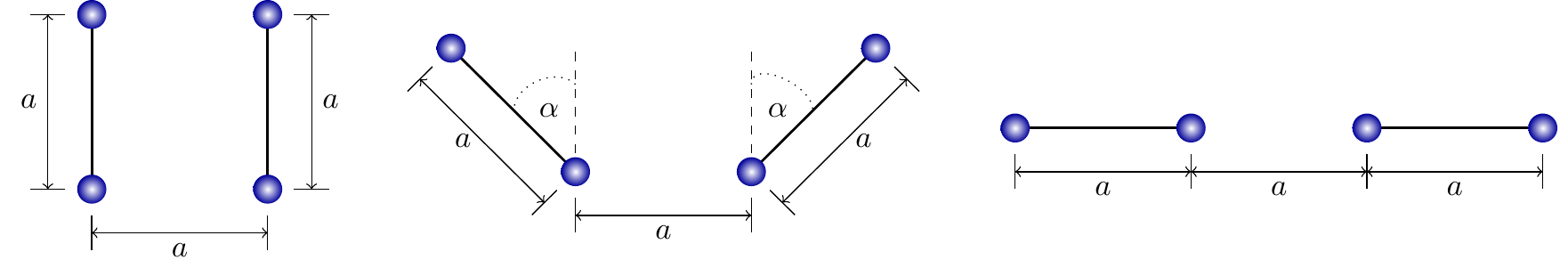}
    \caption{Nuclear configuration for H$_4$ when under going the transition from a square geometry (left) to linear geometry (right).}
    \label{fig:P4toD4}
\end{figure}

In the previous work~\cite{mcclean2020discontinuous}, each element $\kappa$ always corresponds to a rectangular region, and hence this method will be referred to as the {DG-R} below. In particular, \cite{mcclean2020discontinuous} uses a quasi-1D partitioning which is most suited for systems with an (approximate) linear geometry. To illustrate why a general partitioning strategy is needed, consider a minimal example of an H$_4$ molecule undergoing the transition from square to linear geometry (see Figure \ref{fig:P4toD4} and Section~\ref{sec:P4toD4}).

Considering the system to be in linear configuration ($\alpha = 90$), we obtain four {DG-R} elements, but as $\alpha$ tends to zero, we transition to the square configuration and obtain merely two {DG-R} elements containing two atoms each.
Therefore it is not possible to apply a quasi-1D rectangular partitioning strategy while maintaining that each atom belongs to a distinct element (note however in this example we may partition the system into two elements, so that each element contains two atoms for every configuration).

%
%This shows, that a rectangular partitioning along a particular symmetry axis does not treat the individual atoms at the same level, and therewith violates the initial motivation, namely, that every grid point within a DG element is closest to the corresponding atom.

% \begin{figure}
%     \centering
%     \includegraphics[width = 0.85\textwidth]{Images/Tiled_atoms.png}
%     \caption{}
%     \label{fig:TiledAtomsC2H4}
% \end{figure}

\begin{figure}[h!]
    \centering
    \begin{subfigure}{.5\textwidth}
    \centering
    \includegraphics[width=\textwidth]{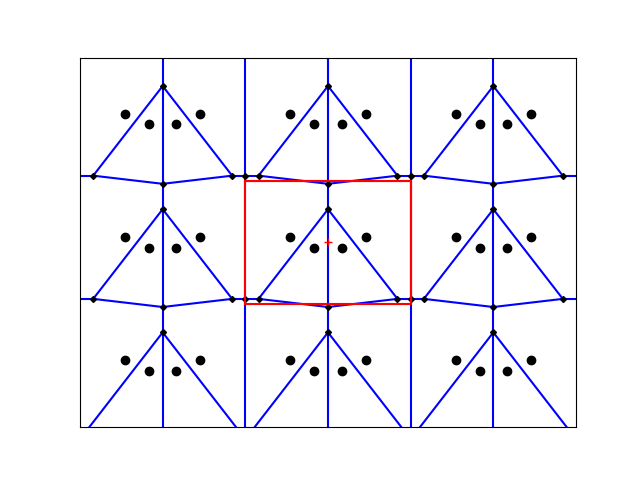}
    \caption{~}
    \label{fig:TiledAtoms}
    \end{subfigure}%
    \begin{subfigure}{.5\textwidth}
    \centering
    \includegraphics[width=1.1\textwidth]{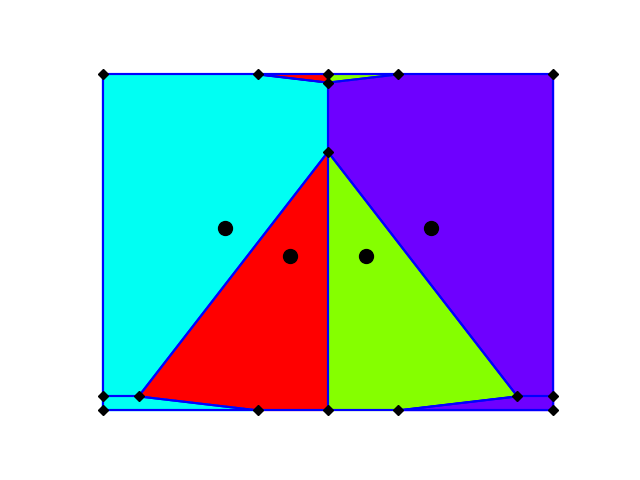}
    \caption{~}
    \label{fig:SupercellVoronoi}
\end{subfigure}
\caption{ (a) Two dimensional tiled supercell. The Voronoi vertices are depicted by black diamonds and the edges connecting the vertices are colored in blue. The red box corresponds to the supercell and hydrogen atoms are depicted by black circles. The system parameter $\alpha = 60$ degrees and as quasi-2D system it is tiled in two dimensions. (b) The corresponding Voronoi decomposition of the supercell. The individual Voronoi cells are the areas in the supercell of identical color.}
\label{fig:Voronoi}
\end{figure}

%\LL{ move figure 2 here} 
%
In order to find a  general partitioning strategy that can be adapted to more complex geometries, we employ the Voronoi decomposition of the considered spatial domain. 
The Voronoi decomposition~\cite{aurenhammer2013voronoi} partitions the domain into a set of Voronoi cells, i.e. a set of disjoint and domain filling subsets, see e.g. Figure~\ref{fig:Voronoi}(b) (this is in fact the Voronoi partitioning defined via periodic distances). 
%
%The Voronoi decomposition can be defined at the level of metric spaces~\cite{aurenhammer2013voronoi}, however, we limit the following definition to the Euclidean space since this is sufficient for our purposes.
%
Consider the supercell $\Omega$ with $M$ nuclei placed at $\vR_1,...,\vR_M$.
The Voronoi cell $\kappa_I$---or Thiessen polygon---associated with the $I$-th nuclear position $\vR_I$ is the set of all points in the supercell $\Omega$ whose distance to $\vR_I$ is not greater than the distance to any other nucleus positioned at $\vR_J$, where $J \neq I$, i.e.
\begin{equation}
\label{eq:Vcell}
\kappa_{I}=\{\vr\in \Omega~|~ \Vert \vr -\vR_{I}\Vert \leq \Vert \vr -\vR_{J}\Vert \;\forall \;J\neq I\}~. 
\end{equation}

While the molecular configuration is always in three-dimensional space, in order to improve efficiency, we may also project all grid points, including the atomic positions to a $d$-dimensional subspace (e.g. a line with $d=1$, or a plane with $d=2$) before applying the partitioning.   
Since we furthermore apply the Voronoi decomposition in a supercells model with periodic boundary conditions, we utilize a periodic Voronoi decomposition.
To that end, we tile $3^d-1$ copies of the original supercell around the original supercell $\Omega$ (see Figure~\ref{fig:Voronoi}(b) for an example with $d=2$).
Characterizing the system of tilled supercells by means of displacement vectors $\{\mc{R}_k\}_{k=1}^{3^d}\subset \mathbb{L}$, we can adjust the definition of the Voronoi cell in Eq.~\eqref{eq:Vcell} in the following way
\begin{equation}
\kappa_{I}=\left\lbrace \vr\in \Omega~\Big|~ \exists \ell ~{\rm s.t. } ~\Vert \vr -(\vR_{I}+\mc{R}_{\ell})\Vert \leq \Vert \vr - ( \vR_{J}+\mc{R}_k)\Vert \;\forall \;J\neq I, k=1,...,3^d\right\rbrace,
\label{eq:Vcell_periodic}
\end{equation}
which is referred to as the periodic Voronoi decomposition (or simply as the Voronoi decomposition below).  
We see that the Voronoi decomposition fulfills by definition the premise that each grid point associated with a {DG-V} element is (periodically) closest to the corresponding atom.  Eq.~\eqref{eq:Vcell_periodic} also provides a practical way for implementing the Voronoi decomposition via a sorting operation. In practice we find that computing the Voronoi decomposition only takes a very small amount of time compared to other components, such as forming the two-electron repulsion integral (ERI) tensor.

%In the procedure above, we 

% %There are several ways to implement a Voronoi decomposition for a given supercell. %
% Regarding the practical implementation used in this study:
% %
% As the Voronoi decomposition for large systems is not the computational bottle neck of the presented calculations, we opt for a `vanilla version' that is numerically very stable. % at a slight loss of computational efficiency. 
% %
% Put differently, we compute the Voronoi decomposition by performing a na\"ive minimization over all grid points while vectorizing as many operations as possible. 
% %
% To that end, we compute the periodic-distance vectors of the system $\{{\bf d}^{(i)}\}_{i=1}^M$, i.e.
% \[
% d_j^{(i)} = \underset{{\bf R}_k}{{\rm min}} \{ \Vert {\bf r}_j-({\bf R}_k + {\bf m}_i) \Vert \},
% \]
% where $i=1,...,M$ and ${\bf r}_j$ are the grid points, and form an auxiliary matrix containing $\{{\bf d}^{(i)}\}_{i=1}^M$ as columns.
% %
% We then use a sorting algorithm, in this case quicksort, along the rows to obtain the different Voronoi cells.    

\subsection{Controlling the number of basis functions}
\label{sec:control}

\begin{figure}[h!]
    \centering
    \includegraphics[width=.75\textwidth]{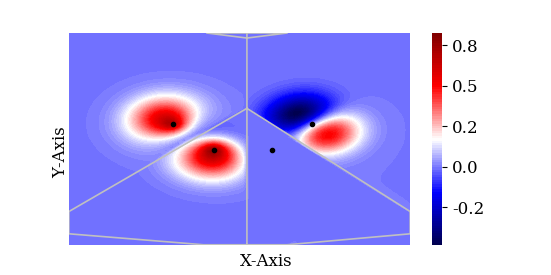} 
    %DGV-func_abs.png for absolute value 
\caption{Three {DG-V} functions for H$_4$ with a system parameter $\alpha = 45$ degrees (see Figure~\ref{fig:P4toD4}) in the X--Y plane, represented in real space; the nuclei positions are depicted by the black dots. The color intensity represents the amplitude $\phi_{\kappa j}$. Each {DG-V} function is localized to its Voronoi element, where the element divisions are shown in gray lines. The {DG-V} functions are represented by linear combination of planewave dual basis functions. Each {DG-V} function is strictly still a continuous function, but its nodal values (defined according to the center of each planewave dual function) are supported within one {DG-V} element only.}
\label{fig:DGV-funcations}
\end{figure}

Since the DG procedure effectively divides one continuous basis functions into several functions that are close to be discontinuous (see Figure~\ref{fig:DGV-funcations}), the number of DG functions will in general be larger than the total number of active space basis functions. 
Note that the number of DG basis functions is connected to the low-rank approximation imposed in Eq.~\eqref{eq:low-rank-phi}.
For systems with increasing sizes, e.g. hydrogen chains and graphene, we control the number of DG basis functions consistently through a relative-tolerance criterion: 
Given an exact singular value decomposition of the basis projection matrix $\Phi_\kappa$ with singular values $\sigma_{\kappa, 1},...,\sigma_{\kappa, N_a}$ ordered non-increasingly, we neglect all singular values that are below a defined threshold $\varepsilon_{\rm tol}$ when normalized by the largest singular value $\sigma_{\kappa, 1}$, i.e. we neglect $\sigma_{\kappa, i}$ if $\sigma_{\kappa, i}/\sigma_{\kappa, 1}  < \varepsilon_{\rm tol}$.
Asymptotically, we find that the number of basis functions per element reaches a constant as the system size increases (see Section~\ref{sec:HChains}). This is also consistent with the studies of the DG basis functions in the context of DFT calculations \cite{lin2012adaptive,HuLinYang2015a}.
However, for small molecules, it is more convenient to  directly control the number of DG basis functions per element, i.e. for a given {DG-V} partitioning, we restrict the number of DG basis functions to an \emph{a priori} defined percentage of the number of active space basis functions.
%

%

% \LL{What are the inputs. Nodal values of AO / orthogoonalized AO basis.  Whether to orthogonalize / how to orthogonalize. Will mention how this is done in each example.
% 1) H4: Original Gaussian basis set.
% 2) H-chain: orthgonalized. In the main text, we talk about the basis obtained from Lowdin orthgonalized Gaussian. only talk about the number of 2ERIs. In the appendix, we provide lambda values (with Lowdin-orthogonalized / SVD-orthogonalized), and also talk about the number of 2ERIs in the SVD-orthogonalized).   
% 3) methane: Just AO without orthogonalization.
% 4) Graphene: Just AO without orthogonalization. (double check).}

We also remark that the SVD truncation criterion may also be affected by the representation of the active space basis functions as the input to the DG procedure. For instance, for a fixed given tolerance $\varepsilon_{\rm tol}$, the DG basis functions may depend on whether the active space basis functions are orthonormalized, or how they are orthonormalized.  We will discuss these in Section~\ref{sec:HChains}.

% of the 
% For the simulations presented here, different choices for the active space basis are possible. 
% %
% Subsequently, we use primarily Gaussian type atomic orbitals, in particular, we will use basis sets that are specifically designed to capture electron-correlation effects.
% %
% Note that as highlighted in Sec.~\ref{sec:DG-general} (after Eq.~\ref{eq:low-rank-phi}), this choice of the active space basis is not orthogonal. 
% %
% Although this does not affect the following energy calculations, as we also aim to compare cost factors for quantum simulations of chemistry, we need to impose a L\"owdin orthonormalization for said comparison. 

\section{Numerical Results}

%\LL{ Names: Discontinuous Galerkin basis with Voronoi partitioning( DG-Vornoi, or DG-V for short). The previous method is called the  Discontinuous Galerkin basis with rectangular partitioning, {DG-R}, or {DG-R}} 

In this section we present in silico applications of the {DG-V} formalism.
To that end, we have implemented the {DG-V} procedure in the PySCF package (version 1.6.4)~\cite{sun2018pyscf} and apply it to the following test systems:

In Section~\ref{sec:P4toD4}, we investigate the flexibility of the Voronoi partitioning with respect to the molecular geometry for the H$_4$ molecule. We study the convergence with respect to the grid size of the underlying planewave dual basis set, as well as the small ``egg-box'' effect due to the use of the grid. We present energy computations at the mean-field, and correlated level of theory comparing the {DG-V} formalism with active space basis set calculations. Furthermore, we demonstrate that even for such a small molecule, the GTO basis sets can become nearly linearly dependent, especially when diffuse basis functions are added. In comparison, the DG basis set is an orthonormal set by construction, and therefore does not suffer from the ill-conditioning issue of the overlapping matrix.

In Section~\ref{sec:HChains}, we investigate the potential benefit of using the {DG-V} procedure for reducing the cost of quantum simulation of chemistry. Compared to GTOs, the use of the {DG-V} basis set can already lead to a smaller number of nonzero two-electron repulsion integrals for a small hydrogen chain consisting of tens of atoms. Furthermore, the number of two-electron integral elements is five orders of magnitude less compared to that of the underlying planewave dual basis set.

% the DG approach becomes advantageous compared to Gaussian for a relatively small system size as a function of the system's size.
% %
% As outlined above, in order to compare cost factors that are influenced by the number of non-zero two electron repulsion integrals, we need to impose a L\"owdin orthonormalization, which is commonly applied to Gaussian type atomic orbitals in the context of quantum simulation of chemistry.
% %
% %and the $\lambda$-value, i.e.
% %\[
% %\lambda = \sum_{\kappa, \kappa'; p,q,r,s} |v_{\kappa, \kappa'; p,q,r,s}|.
% %\]
% %
% %Both quantities are relevant for quantum simulations based on the \emph{linear combination of unitaries} approach. 
% %
% A set of rather simple systems that allow large sampling of said scaling properties is given by quasi one-dimensional hydrogen chains. 
% %
% We compare the {DG-V} results with the results presented in Ref.~\cite{mcclean2020discontinuous} on the {DG-R} procedure, and present new data giving insight on the influence of L\"owdin orthonormalzation to the scaling behavior. 

%
In Section~\ref{sec:Methane}, we extend the numerical investigations of \emph{ab initio} quantum-chemical simulations to methane with a  three dimensional Voronoi partitioning.
We present energy computations at the mean-field, and correlated level of theory comparing the {DG-V} formalism with active space basis set calculations. 

In Section~\ref{sec:Graphene}, we apply the {DG-V} formalism to graphene to investigate the performance for a crystalline system. We study the accuracy of Hartree-Fock calculations and MP2 calculations with {DG-V} basis for the supercells of 4 atoms, 8 atoms, and 16 atoms.  %\LL{ add} 

Additional numerical results are presented in the supplementary material. 

%
%\LL{Truncation criterion: In all cases, we need to have the number of basis functions per atom. In all cases, our total number of basis is larger than that of the AO basis, but asymptotically this is a constant times the AO (shown in the h-chain example). Because of the small system size, we may directly control the number of basis functions. The number of {DG-V} basis per element is xx percentage of the total number of AOs. In the thermodynamic limit, the percentage should go to zero. For larger systems, for consistency we set a relative SVD tolerance to control the number of {DG-V} basis functions systematically (define this somewhere). 1) H4: basis number. 2) H-chain: orthgonalized. SVD relative tolerance. We also report the number of basis functions. 3) methane: basis number. 4) Graphene: relative tolerance.} 

\subsection{H$_4$ transition from square to linear geometry}
\label{sec:P4toD4}
In order to investigate the flexibility of the {DG-V} procedure when applying it to molecular systems, we seek a model system that allows for a continuous variation of the degree of quasidegeneracy, while being small enough to enable computations using various and large basis sets at an affordable cost.
Moreover, we seek a molecular geometry that cannot be covered by means of a rectangular partitioning procedure. 
A minimal system that fits these requirements is H$_4$ transitioning from square to linear geometry, see Figure~\ref{fig:P4toD4}. 
The degree of quasi degeneracy is here steerable thought the parameter $\alpha$; as $\alpha$ tends to zero the systems becomes metallic, i.e. the HOMO-LUMO gap vanishes.
Following Ref.~\cite{jankowski1980applicability}, we set $a = 2.0$ (a.u.).

Our selection of the active space basis set includes Pople basis sets w.o./w. diffuse basis functions (3-21G, 3-21++G, 6-31G, 6-31++G, 6-311G, and 6-311++G), as well as a correlation consistent basis set w.o./w. diffuse basis functions (cc-pVDZ, and aug-cc-pVDZ).
These basis sets are not orthogonal, while the {DG-V} basis functions will be orthonormal as outlined in Section~\ref{sec:DG-general}. In order to investigate the sensitivity of the Hartree--Fock energy with respect to the truncation of the {DG-V} basis set per element, we apply a truncation threshold relative to the maximal number of {DG-V} basis functions per element, which is equal to the total number of active space basis functions. 
This corresponds to \emph{a  priori} restricting the number of {DG-V} basis functions to be a fixed percentage of the number of active space basis functions, as outlines in Section~\ref{sec:control}.

\begin{figure}[h!]
    \centering
    \begin{subfigure}{.33\textwidth}
    \centering
    \includegraphics[width=1.\textwidth]{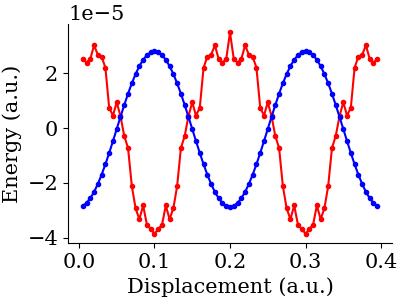}
    \caption{~}
    \end{subfigure}%
    \begin{subfigure}{.33\textwidth}
    \centering
    \includegraphics[width=1.\textwidth]{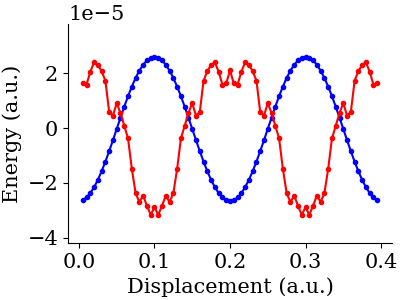}
    \caption{~}
\end{subfigure}
\begin{subfigure}{.33\textwidth}
    \centering
    \includegraphics[width=1.\textwidth]{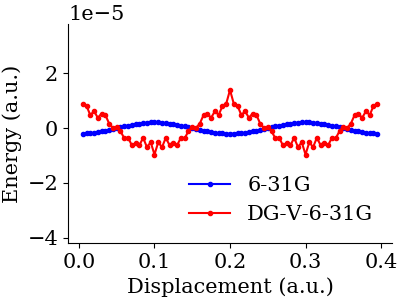}
    \caption{~}
\end{subfigure}
\caption{
``Egg-box" effect for H$_4$ with $\alpha = 60$ degree, 6-31G as active space basis set and an underlying real-space grid with grid spacing of $0.2$ (a.u.). Different energy contributions (a) total energy (b) mean field energy (c) MP2 correlation energy are plotted as functions of the displacement along the horizontal symmetry axis (cf Figure~\ref{fig:P4toD4}). 
%\wxj{can you add x-labels and y-labels, including units?} 
}
\label{fig:EggBox631g}
\end{figure}

In the following calculations we center the molecule in a supercell of size $12\times 12 \times 12$ (a.u.).
As the {DG-V} formalism is based on a real-space grid discretization, we first study the convergence with respect to the grid spacing, i.e. in order to prohibit errors through the use of a too coarse grid, we determine the utilized grid spacing based on the convergence of the Hartree--Fock energy with respect to the underlying grid spacing, see e.g. Figure~\ref{fig:Grid_conv_6311g} for an example using 6-311G as active space basis showing the convergence of the mean-field energy, as well as convergence of the correlation energies at the MP2 level of theory. 
These calculations show that the Hartree--Fock energy is sufficiently converged for a grid spacing of $0.2$ (a.u.), which we shall use subsequently.
Another way to study the convergence of the grid spacing is to investigate the ``egg-box'' effect, i.e. the unphysical phenomenon that the energies are affected by the positions of the grid points relative to the nucleus. 
%\wxj{what does this mean?}
%
We compare the egg-box effect in total, mean-field and correlation energy (at the MP2 level) of the {DG-V} formalism and the periodic boundary condition module of PySCF, see Figure~\ref{fig:EggBox631g} for 6-31G as active space basis and Figure~\ref{fig:EggBoxccpvdz} for cc-pVDZ as active space basis. We find that the egg-box effect caused by the {DG-V} procedure is on the order of $3\times 10^{-5}$ Hartree and is negligible. The size of the error is also comparable to that in the periodic boundary condition module of PySCF.

% \begin{figure}
%     \centering
%     \includegraphics[width=\textwidth]{Images/EggBox_631g_fine1.png}
%     \caption{``Egg-box" effect for H$_4$ with $\alpha = 60$ degree with 6-31G as active space basis set and an underlying real-space grid with grid spacing of $0.2$ (a.u.). Different energy contributions are plotted as function of the displacement along the horizontal symmetry axis (cf Figure~\ref{fig:P4toD4}). {\color{red}Xiaojie: should we use the same style like (a), (b), (c) here}}
%     \label{fig:EggBox631g}
% \end{figure}

% Preliminary calculations have moreover shown that the utilized supercell size is sufficient in order to prohibit bonding over the boundaries of the supercell, i.e. the vacuum package is sufficiently large.   
%
In order to numerically investigate the applicability of the {DG-V} procedure to a metal--insulator type transition, we gradually change the system parameter $\alpha$, i.e. we perform the following energy calculations at $\alpha = 15,\,30,\,45,\,60,\,75,\,90$ degrees. 
%
%We observe that the {DG-V} elements change as a function of system parameter $\alpha$, always providing the ideal partitioning of the supercell.
%
%In the transition process of H$_4$ from P4- to D4-configuration the {DG-V} elements change as a functions of the parameter $\alpha$.
%
%Since a rectangular partitioning of the supercell, as presented in Ref.~\cite{mcclean2020discontinuous}, is rather limiting to the system's geometry, we use a periodic Voronoi partitioning order to compute quasi two-dimensional---and later on three dimensional--systems.
%
%The periodicity of the Voronoi decomposition is achieved by tiling copies of the considered supercell around the original supercell, see Figure~\ref{fig:Voronoi}(a) for a two dimensional example.
%
%Performing a minimization procedure (as described in Sec.~\ref{sec:voronoi}) over the grids discretization while considering all atoms of the tiled system yields the periodic Voronoi decomposition. 
%
%An example of such a periodic Voronoi decomposition is depicted in Figure~\ref{fig:Voronoi}(b) together with the nuclear positions and the Voronoi vertices. 
%
%
We first note that, compared to the rectangular partitioning of the system, the {DG-V} procedure provides sufficient freedom to decompose the molecular system in an optimal way, i.e. grouping the grid points that are periodically closest to a particular atom into one Voronoi cell. 
By construction, the {DG-V} elements change accordingly as we change the parameter $\alpha$, so that every atom is always located in the interior of each element. 

\begin{figure}
    \centering
    \includegraphics[width=0.5\textwidth]{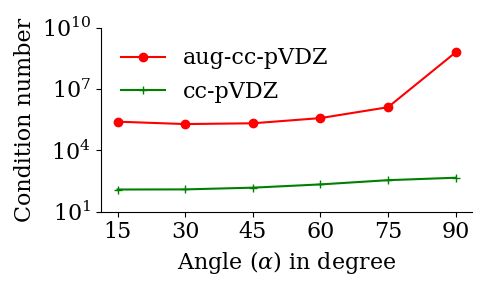}
    \caption{Condition number of the overlap matrix as a function of the parameter $\alpha$ for (aug-)-cc-pVDZ basis discretization.}
    \label{fig:ConditionNumber}
\end{figure}

The appearance of (near-computational) linear dependence in quantum-chemical simulations is a common problem when using Gaussian type atomic orbitals.
This is indeed the case for molecular systems, and even more so for crystalline systems. The case is particularly severe when diffuse functions are added to the basis set, which are important for describing, e.g., anions or dipole moments~\cite{helgaker2014molecular}.
%
%See Figure~\ref{fig:ConditionNumber} for an example of the overlap matrix' condition number when using the (aug)-cc-pVDZ basis. 
%
The DG approach remedies this linear dependence while extending the descriptive power of the basis set. 
In order to monitor the linear dependence of a given basis set, we compute the condition number of the overlap matrix.
When we add diffuse basis functions the condition number of the overlap matrix is amplified, 
%\wxj{the condition number of the overlapping matrix is amplified}
see Figure~\ref{fig:ConditionNumber} for an example using (aug-)-cc-pVDZ and Figure~\ref{fig:ConditionNumberDiffBasis} for a comparison across different basis sets.
This is an expected behaviour as the diffuse functions are extended Gaussian basis functions with small exponents, however, we highlight again that by construction, the condition number of the {DG-V} overlap matrix is equal to one, and is therefore \emph{always} well-conditioned.

\begin{figure}[h!]
    \centering
    \begin{subfigure}{.49\textwidth}
    \centering
    \includegraphics[width=1.\textwidth]{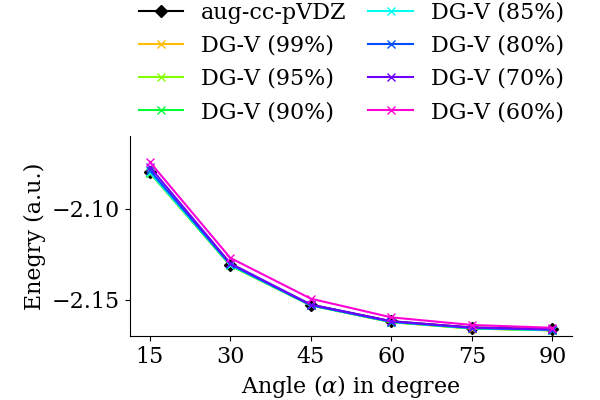}
    \caption{~}
    \end{subfigure}%
    \begin{subfigure}{.49\textwidth}
    \centering
    \includegraphics[width=1.\textwidth]{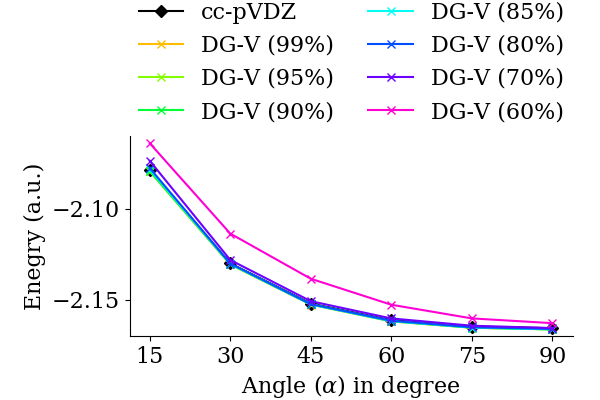}
    \caption{~}
\end{subfigure}
\begin{subfigure}{.49\textwidth}
    \centering
    \includegraphics[width=1.\textwidth]{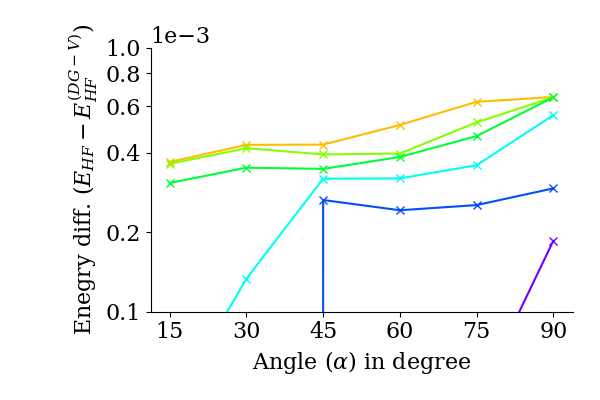}
    \caption{~}
\end{subfigure}
\begin{subfigure}{.49\textwidth}
    \centering
    \includegraphics[width=1.\textwidth]{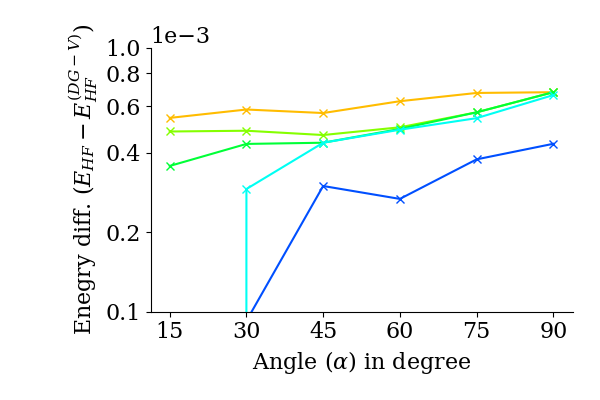}
    \caption{~}
\end{subfigure}
\caption{
Subplots (a) and (b) show the absolute energies for (aug)-cc-pVDZ for different truncations of the number of {DG-V} basis functions. The relative numbers of {DG-V} basis functions kept per element are reported in percentage. The {DG-V} solutions are compared to the solution provided by the periodic boundary condition module in PySCF ( depicted by the black dimond markers). The subplots (c) and (d) show the differences between the PySCF solution using the (aug)-cc-pVDZ basis set and the {DG-V} solutions, i.e. $ E_{\rm HF} - E_{\rm HF}^{\rm ({DG-V})}$ on a semilogarithmic scale. Note that as the number of the {DG-V} basis functions increases, the energy from the (aug)-cc-pVDZ basis set is higher than that from the {DG-V} basis set.}
\label{fig:PES_mf_ccpvdz}
\end{figure}

As outlined above, the transition of H$_4$ from square to linear geometry furthermore exemplifies the flexibility of the {DG-V} approach applied to quasi two-dimensional molecular systems, describing configurations with small HOMO-LUMO gap as well as configurations with well-separated HOMO and LUMO (as $\alpha$ tends to 90 degrees). 
In the following we investigate the robustness of the {DG-V} approach with respect to the imposed truncation along the change of the molecular configuration.
%
%To that end, we compute the potential energy surface as function of the model parameter $\alpha$ for different truncation levels at the mean-field level of theory.
%
See Figure~\ref{fig:PES_mf_ccpvdz} for an example using (aug-)cc-pVDZ and Figure~\ref{fig:MFtruncationApp321}, \ref{fig:MFtruncationApp631}, and \ref{fig:MFtruncationApp6311} for a comparison across different basis sets.
Clearly, the {DG-V} basis set is---in its untruncated form---larger than the active-space basis, and yields therefore lower energies (given that the primitive basis set is of sufficient size).
%
%In order to reduce the overall number of basis functions, we truncate the VdG basis sets within the individual Voronoi cells.
%
As we are interested in reducing the number of basis functions per Voronoi element, we used a truncation of the {DG-V} basis functions per element relative to the maximal number of {DG-V} basis functions, i.e. only keeping a certain percentage of the number of {DG-V} basis functions per Voronoi element. 
We observe that the Hartree--Fock energy of the considered system is fairly robust with respect to said truncation, see Figure~\ref{fig:PES_mf_ccpvdz}.
For clarity, we list the minimal and maximal difference in millihartree (i.e. max and min of $E_{\rm HF} - E_{\rm HF}^{\rm (DG)}$) along the PES between the Hartree--Fock energy in active-basis representation and {DG-V} discretization for different level of truncation in Table~\ref{tab:MinimalDifferenceHF} and~\ref{tab:MaximalDifferenceHF}. 

\begin{figure}[h!]
    \centering
    \begin{subfigure}{.49\textwidth}
    \centering
    \includegraphics[width=1.\textwidth]{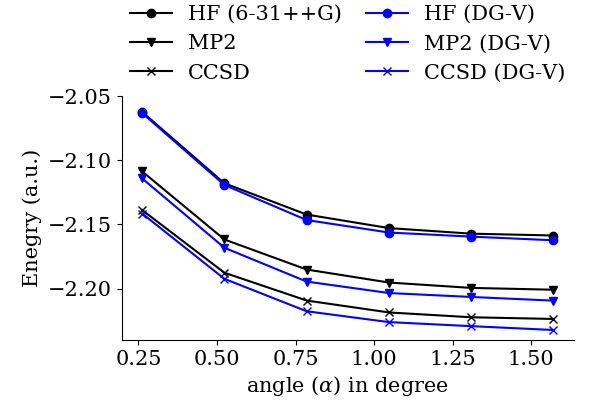}
    \caption{~}
    \end{subfigure}%
    \begin{subfigure}{.49\textwidth}
    \centering
    \includegraphics[width=1.\textwidth]{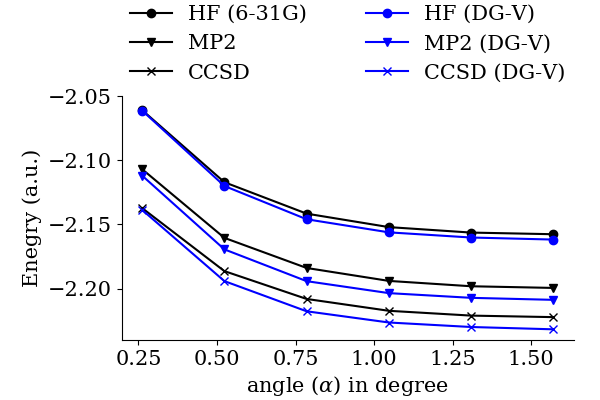}
    \caption{~}
\end{subfigure}
\begin{subfigure}{.49\textwidth}
    \centering
    \includegraphics[width=1.\textwidth]{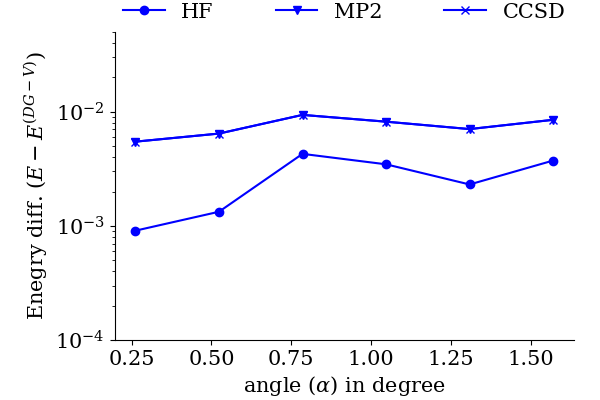}
    \caption{~}
\end{subfigure}
\begin{subfigure}{.49\textwidth}
    \centering
    \includegraphics[width=1.\textwidth]{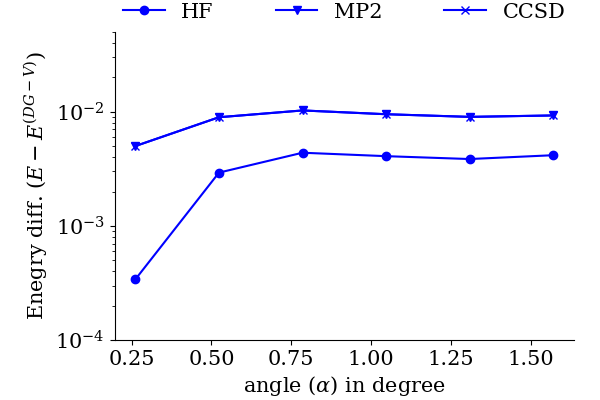}
    \caption{~}
\end{subfigure}
\caption{
Subplots (a) and (b) show the absolute energies for mean field, MP2 and CCSD energy computations. The truncations of the number of {DG-V} basis functions are extracted from previous mean field computations and was \emph{a priori} set to be 60\% and 70\% for 6-31++G and 6-31G, respectively. The {DG-V} solutions are compared to the solution provided by the periodic boundary condition module in PySCF (depicted by the black curves).
The subplots (c) and (d) show the differences between the PySCF solution and the {DG-V} solutions, i.e. $ E_{\rm HF} - E_{\rm HF}^{\rm ({DG-V})}$ on a semilogarithmic scale. }
\label{fig:correlation631G}
\end{figure}

We furthermore investigate the performance of {DG-V} including electron-correlation effects at the level of MP2 and CCSD. 
Based on Table~\ref{tab:MinimalDifferenceHF}, we can set an appropriate level of truncation.
The corresponding potential energy surfaces are in presented in  Figure~\ref{fig:correlation631G} for the Pople bases 6-31G and 6-31++G, see also Figure~\ref{fig:correlation321G}, \ref{fig:correlation6311G} and \ref{fig:correlationccpvdz} for a comparison across different basis sets. %
The ratio of {DG-V} basis functions per Voronoi cell to the number of basis functions in regular discretization together with the respective level of truncation is listed in Table~\ref{tab:NumberOfBasisFunctions}.
It is interesting to observe that while the mean-field energy obtained from the {DG-V} basis set is only slightly lower than that from the GTOs, the correlation energy from the {DG-V} basis set is considerably lower. This indicates that the {DG-V} basis set can be effective for capturing  dynamical correlation effects.

% \begin{figure}
%     \centering
%     \includegraphics[width=0.8\textwidth]{Images/Corr631G.png}
%     \caption{Comparison of the {DG-V} and regular basis discretization for correlated calculations. The level of truncation for the {DG-V} basis functions per element are 60\% and 70\% for 6-31++G and 6-31G, respectively.\wxj{should we use the same figure style (a), (b), (c).}}
%     \label{fig:correlation631G}
% \end{figure}

\begin{center}
\begin{table}[ht]
\centering
\begin{tabular}{r|cccc}%{@{}*{7}{l}}
\br
Basis sets &  aug-cc-pVDZ & cc-pVDZ & 6-311++G & 6-311G \\
\mr
No. AO & 36   & 20   & 16   & 12\\
No. {DG-V} AO   & 31   & 18   & 12   & 9 \\
\mr
{DG-V} trunc.   & 85\% & 90\% & 70\% & 70\% \\
\mr
\mr
Basis sets &  6-31++G & 6-31G & 3-21++G & 3-21G\\
\mr
No. AO & 12   & 8    & 12   & 8 \\
No. {DG-V} AO &  8   & 6    &  9   & 7 \\
\mr
{DG-V} trunc.   & 60\% & 70\% & 70\% & 80\% \\
\br
\end{tabular}
\caption{Absolute number of {DG-V} basis functions per Voronoi element and the total number of active space basis functions for various basis sets. The corresponding truncation threshold for the number of {DG-V} basis functions is listed in percentage ({DG-V} trunc.) and is to be understood relative to the maximal number of {DG-V} basis functions per element (i.e. the number of active space basis functions).}
\end{table}
\label{tab:NumberOfBasisFunctions}
\end{center}

% We furthermore investigate the egg-box effect where the Voronoi decomposition is fixed, see Figure~\ref{fig:EggBox_fix_321g} for 3-21G as active space basis.
% %and Figure~\ref{?}--\ref{?} for a comparison across different basis sets.
% %
% \begin{figure}
%     \centering
%     \includegraphics[width=\textwidth]{Images/EggBox_321g_fix1.png}
%     \caption{\FF{Egg box effect for H$_4$ with $\alpha = 90$ degree (linear geometry) with 3-21G as active space basis set and an underlying real-space grid with grid spacing of $0.2$ (a.u.). Different energies are plotted as function of the displacement along the horizontal symmetry axis (cf Figure~\ref{fig:P4toD4}).
%     For a fixed Voronoi decomposition (which here coincides with the rectangular partitioning of the domain as the system shows a linear geometry), the egg box effect overlays with an error caused by using a subotimal decomposition of the system.}\LL{ Which version of DG is this? Uniform partition? Need to label it clearly} }
%     \label{fig:EggBox_fix_321g}
% \end{figure}

\subsection{Hydrogen chains}
\label{sec:HChains}
Next, we investigate hydrogen chains of various lengths as presented in Ref.~\cite{mcclean2020discontinuous}, aiming at empirical data for cost factors in quantum simulation of chemistry.
This quasi one-dimensional system is well suited for examining different scaling behavior of the {DG-V} formalism as a function of the system size.
As we systematically increase the number of atoms, we choose a consistent truncation procedure of the {DG-V} basis functions by means of the absolute size of the singular values. 
In the particular case of hydrogen chains, we observe that around 90\% of singular values are enclosed in the interval $[0,1]$, and the largest singular value is approximately $10$. 
Hence, in this special case, the truncation of the singular values with respect to the absolute size, can be straightforwardly related to a truncation criterion with respect to relative size. 
%
%Given our experience with the truncation criterion of the {DG-V} procedure, we recommend a truncation with respect to the relative size of the singular values or a direct control of the basis set size, as presented for H$_4$ and CH$_4$. \LL{ i do not understand the meaning of this sentence and how it is related to what was said earlier.} 

%

%
Aligning with the previous simulations we here use the cc-pVDZ basis. 
Furthermore, as we aim at a cost factor comparison for quantum simulation of chemistry, we need to construct a second quantized Hamiltonian with an orthogonormal basis set. To this end we compare the performance of the {DG-V} basis with the L\"owdin orthonormalized~\cite{Loewdin1950} cc-pVDZ atomic orbital basis.   

\begin{figure}
    \centering
    \includegraphics[width = 0.7\textwidth]{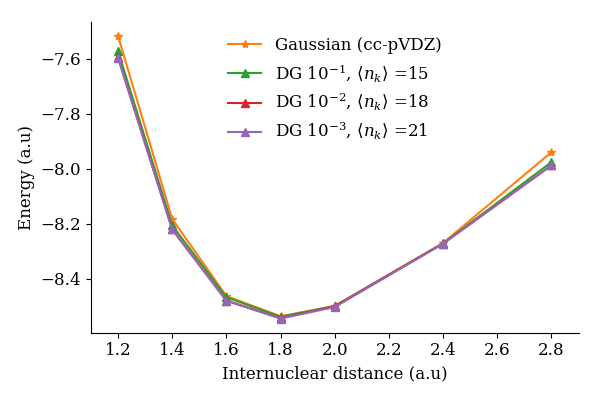}
    \caption{Potential energy surfaces for H16 in {DG-V} basis with different truncation tolerances. The average number of {DG-V} basis per atom along the PES is given by $\langle n_\kappa \rangle$ for each {DG-V} fit tolerance, i.e. the SVD cutoff threshold. For comparison the number of primitive functions per atom here is approximately 52000 . The primitive basis set is more expressive by design than the active-space Gaussian cc-pVDZ basis against which the {DG-V} fit is performed. This allows even fairly loose {DG-V} fits to match the accuracy of the active-space basis. }
    \label{fig:pesH16}
\end{figure}

\begin{figure}
    \centering
    \includegraphics[width = 0.7\textwidth]{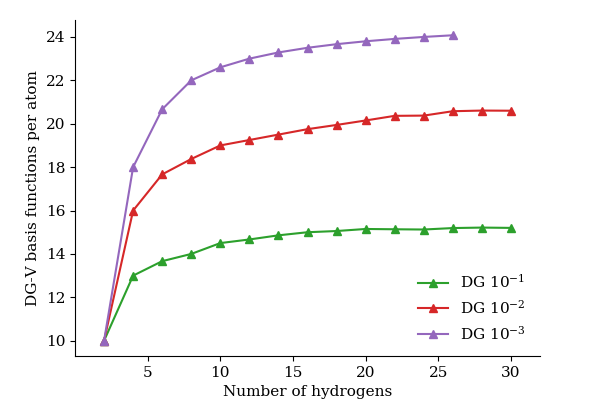}
    \caption{Convergence of block size $n_k$ as a function of system size. The average number of {DG-V} basis functions per atom at $1.4$ (a.u.) SVD tolerances of $10^{-1}$, $10^{-2}$ and $10^{-3}$. For a fixed accuracy, it is observed that the average number of {DG-V} functions per Voronoi call converges as a function of system size.  }
    \label{fig:AOs_per_vdg_elem}
\end{figure}

We also present numerical evidence that the number of {DG-V} basis functions per Voronoi cell as function of the system size converges to a constant that depends on the truncation tolerance. 
%
%We present computational values ranging from H$_4$ to H$_{30}$. 
%
%To that end, we investigate hydrogen chains of increasing length in a Gaussian cc-pVDZ basis (the active space to be fit) and the planewave dual functions (the primitive basis set) with refinement built to match the accuracy of the Gaussian basis set to a specified tolerance at the mean-filed level of theory.
%
The hydrogen chains H$_n$ with $n-1$ bond length of $1.7 $ (a.u.) are centered in a supercell of size $(n\cdot 3.6+ 10) \times 10 \times 10$ (a.u.).  
The underlying grid spacing was chosen to be $0.2$ (a.u.).
In the case of H$_{30}$ this yields a super cell of size $118 \times 10 \times 10$ with  $590 \times 50 \times 50$ grid points which corresponds to $1475000$ planewave dual functions, i.e. approximately $49000$ basis functions per atom. The large number of planewave dual functions needed is due to that cc-pVDZ is a basis set designed for all-electron calculations. 
Using this configuration, we find that, even for an aggressive truncation threshold, the result from the {DG-V} basis can reliably matches the accuracy of the results from Gaussian orbitals, while each {DG-V} basis function is far more compact; see Figure~\ref{fig:pesH16} for a comparison of potential energy surfaces (PES) in {DG-V} discretizations of different tolerances with the corresponding PES in the active space basis set.
% %
% Before analyzing the scaling of the nonzero two-electron repulsion integrals, we verify the accuracy of the {DG-V} basis in Figure~\ref{fig:pesH16} by comparing potential energy surfaces (PES) in {DG-V} discretizations of different tolerances with the corresponding PES in the active space basis set. 
% %
% We find that the result from the {DG-V} basis can reliably match the accuracy of the results from Gaussian orbitals, while each {DG-V} basis function is far more compact.
% %
% We here use a grid spacing of $0.2$ (a.u) and obtain accurate results for a rather aggressive truncation threshold of $10^{-1}$ in the {DG-V} basis. \LL{ is the grid spacing consistent with the previous discussion of the number of grid points? if so it should be combined} 
%
In fact, the energies obtained from the {DG-V} basis are slightly lower than those from the Gaussian basis set. 
This is because as $n_\kappa$ increases, the span of the Gaussian basis set becomes approximately a subspace of the span of the {DG-V} basis set, and hence the {DG-V} basis can possibly yield lower energies due to the variational principle.
% 
%Aligning with the results in Ref.~\cite{mcclean2020discontinuous}, we find the main limiting factor in the VdG procedure's accuracy to be the cc-pVDZ basis set to which VdG is fitted. 
%
The results suggest that the overall accuracy of the energy  is relatively insensitive even to rather aggressive singular value truncations in the {DG-V} decomposition procedure, at least for quasi one-dimensional systems.
We also confirm that the average number of {DG-V} basis functions per atom for fixed truncation thresholds in the {DG-V} procedure converges rather rapidly to a constant, see Figure~\ref{fig:AOs_per_vdg_elem}.  

%\LL{ let us put the discussion of the $\lambda$ value and the different orthogonalization strategies in the appendix.} 

\begin{figure}
    \centering
    \includegraphics[width = 0.75\textwidth]{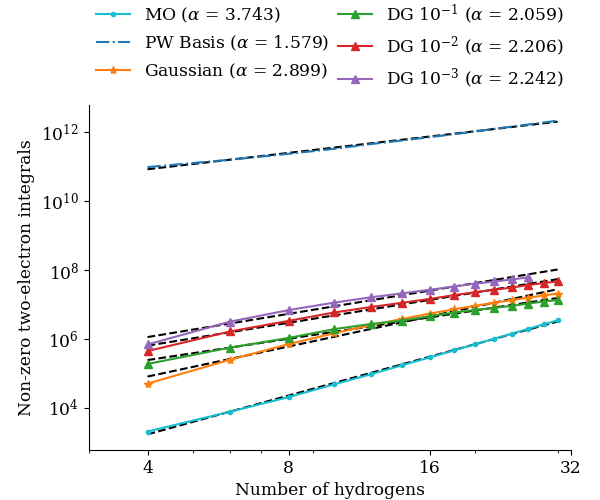}
    \caption{The number of nonzero two-electron integrals in different representations for SVD-truncation tolerances of $10^{-1}$, $10^{-2}$ and $10^{-3}$, plotted on a log–log scale. We fit a trendline plotted with black dots to extract the scaling as a function of system size as $N^{\alpha} + c$ for some constant $c$, and list the exponent $\alpha$ beside each representation in the legend. In the shown graph, we neglect integral contributions smaller that $10^{-6}$.  }
    \label{fig:NNZ-ERI-H-chain}
\end{figure}

With the parameter settings above, we investigate the scaling of the number of nonzero two-electron integrals in Figure~\ref{fig:NNZ-ERI-H-chain}. 
Analogous to the procedure in Ref.~\cite{mcclean2020discontinuous}, we neglect two-electron repulsion integrals with an absolute value smaller than $10^{-6}$. 
In particular, when the SVD truncation tolerance is set to $10^{-1}$, the {DG-V} basis set becomes advantageous when the system size is larger than $14$ atoms. 
Comparing with results from Ref.~\cite{mcclean2020discontinuous}, we find that the scaling of the number of nonzero two-electron integrals with respect to the number of atoms is lower in the present work using the Gaussian basis set, and the value of $\alpha$ is reduced from $3.9$ to $3.0$. %
A more careful investigation reveals that the difference is affected by orthogonalization strategies of the GTOs. 
In this work, we apply the L\"owdin orthonormalized Gaussian orbitals, while Ref.~\cite{mcclean2020discontinuous} uses the SVD-orthonormalized Gaussian orbitals (the number of nonzero two-electron integrals is reproduced in Figure \ref{fig:nnz_eri_l_vs_nl} using the PySCF based code in this work). 
%

% see that the L\"owdin orthonormalization affects the scaling of the two-electron repulsion integrals.
% %

% we notice that the application of L\"owdin orthonormalization affects the observed scaling. 
% %
% Whereas in Ref.~\cite{mcclean2020discontinuous} 
% an arbitrary orthonormalization of the active space basis functions was chose by means of singular value decomposition, we here choose $U$ such that
% \[
% U = \text{argmin} \left\lbrace 
% \Vert \Phi - \tilde{U} \Vert_{\rm F} ~\Big|~  \tilde{U}^T \tilde{U} = Id  \right\rbrace .
% \]

%
%Depending on the truncation threshold imposed in the VdG procedure this crossing appears later.  

\subsection{Methane}
\label{sec:Methane}
We use the CH$_4$ molecule to demonstrate the {DG-V} procedure with a 3D partitioning strategy. 
Similar to the calculation presented in Section~\ref{sec:P4toD4}, we here consider the methane molecule centered in a supercell of size $15 \times 15 \times 15$ (a.u.) using a grid spacing of $0.125$ (a.u.).
We again use Gaussian type atomic orbitals as active space basis set and similar to Section~\ref{sec:P4toD4}, the $\Phi$ matrix is non-orthogonal.
As CH$_4$ is a relatively small system we can control the {DG-V} basis-set size manually by \emph{a priori} defining a truncation threshold for the number of {DG-V} basis functions per element relative to the maximal number of {DG-V} basis functions per element, i.e. the total number of active space basis functions.

\begin{figure}[h!]
    \centering
    \includegraphics[width = 0.6\textwidth]{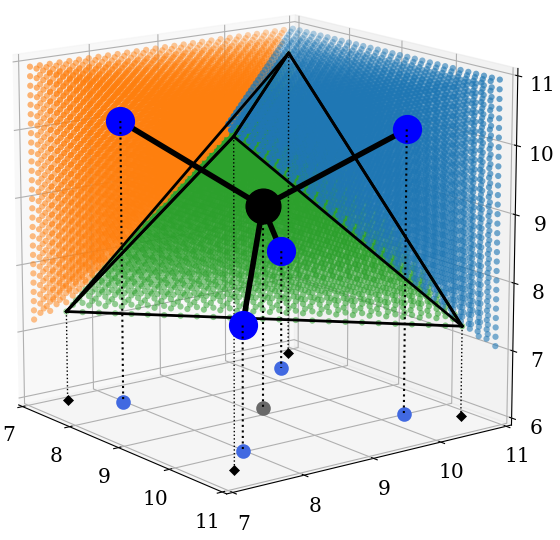}
    \caption{Voronoi decomposition of CH$_4$. The figure shows three {DG-V} element: the green region corresponds to the {DG-V} element of the carbon atom (depicted black), the orange and blue regions correspond to the {DG-V} elements of a hydrogen atom (depicted blue), respectively. The figure furthermore shows the projections of the atoms and the Voronoi vertices onto the X-Y-plane in order to increase the clarity of the picture.}
    \label{fig:ch4_voronoi}
\end{figure}

The application to methane shows again the increased flexibility of the Voronoi decomposition opposed to the previously implemented rectangular decomposition of the considered supercell. 
For a graphical illustration of the decomposition obtained for methane, we show in Figure~\ref{fig:ch4_voronoi} three different {DG-V} elements.
The green-colored grid points correspond to the {DG-V} element around the centered carbon atom (depicted in black). 
The orange- and blue-colored grid points correspond to two different hydrogen atoms (depicted in blue), respectively.

Similar to the calculations performed for H$_4$, we apply the {DG-V} procedure to different atomic orbital sets: 6-31G,  6-31++G,  6-311G, 6-311++G, and the correlation consistent basis set with diffuse basis functions cc-pVDZ, and  aug-cc-pVDZ.
%
%Analogously to the previous molecular calculations, we consider the methane molecule to be placed in a large supercell with a grid spacing of $0.125$ (a.u.).
%
%The exact coordinates and supercell size are given in supplementary material.
%
%We perform an investigation of the mean-field energy and its sensitivity with respect to basis set truncation. 
%
%As we are again investigating the energy sensitivity when reducing the number of basis functions, we perform the basis set truncation with respect to total number of basis functions per element, i.e. we again only consider a percentage of the full {DG-V} basis set. 
%
We consider methane to be in equilibrium geometry and investigate different (diffuse) basis sets. 
The results are listed in Table~\ref{tab:ch4Meanfiled}.

\begin{table}[]
    \centering
    \begin{tabular}{r|c|cccc}
         \br
         Basis set   & AO & {DG-V} (100\%) & {DG-V} (90\%) & {DG-V} (80\%) & {DG-V} (70\%) \\
         \mr
         aug-cc-pVDZ & -7.853 & -7.872 (60) & -7.870 (54) & -7.864 (48) & -7.856 (42)\\
         cc-pVDZ     & -7.851 & -7.864 (34) & -7.864 (31) & -7.847 (28) & -7.837 (24)\\
         6-311++G    & -7.848 & -7.862 (34) & -7.862 (31) & -7.860 (28) & -7.857 (24)\\
         6-311G      & -7.848 & -7.861 (25) & -7.861 (23) & -7.859 (20) & -7.855 (18) \\
         6-31++G     & -7.834 & -7.863 (25) & -7.861 (23) & -7.847 (20) & -7.847 (18) \\
         6-31G       & -7.833 & -7.857 (17) & -7.844 (16) & -7.841 (14) & -7.829 (12)\\
         \br
    \end{tabular}
    \caption{Mean-filed energy calculations for CH$_4$ at equilibrium configuration comparing the periodic boundary condition module in PySCF and the {DG-V} procedure. The energy values are given in Hartree. The truncation threshold for the number of {DG-V} basis functions relative to the maximal number of {DG-V} basis functions per element is reported in percentage (top row in parenthesis); the corresponding absolute number of basis functions is listed in parenthesis behind the individual energies. 
    %\wxj{can you add the units to this table and the following table}
    }
    \label{tab:ch4Meanfiled}
\end{table}

We observe that for all basis sets tested the {DG-V} procedure yielded improved energy results. Moreover, we observe that in most cases, except for cc-pVDZ and 6-31G, an energy improvement is obtained with a {DG-V} basis set truncation as aggressive as 30\%.   
%\wxj{these are two sentences}
%
Based on the accuracy reported in Table~\ref{tab:ch4Meanfiled}, we choose the truncation threshold for the individual active space basis sets and perform correlated calculations at the MP2 level.
The energy results are listed in Table~\ref{tab:ch4MP2}, together with imposed relative truncation threshold and the number {DG-V} basis functions per element.
We see that in all cases, the {DG-V} procedure improves the correlation energies by more than $0.01$ (a.u.), which indicates the effectiveness of the {DG-V} basis functions for capturing dynamical correlation effects.

\begin{table}[]
    \centering
    \begin{tabular}{r|c|c|cc}
         \br
         Basis set   & AO & {DG-V} & {DG-V} trunc.& No. {DG-V} AO\\
         \mr
         aug-cc-pVDZ & -0.168 & -0.184 & 70\% & 42\\
         cc-pVDZ     & -0.162 & -0.181 & 90\% & 31\\
         6-311++G    & -0.104 & -0.123 & 70\% & 24\\
         6-311G      & -0.103 & -0.121 & 70\% & 18\\
         6-31++G     & -0.101 & -0.117 & 70\% & 18\\
         6-31G       & -0.100 & -0.115 & 80\% & 14 \\
         \br
    \end{tabular}
    \caption{MP2 correlation energy for CH$_4$ at equilibrium configuration comparing the periodic boundary condition module in PySCF and the {DG-V} procedure. The energy values are given in Hartree. The truncation threshold relative to the maximal number of {DG-V} basis functions per element is reported in percentage ({DG-V} trunc.) together with the corresponding absolute number of {DG-V} basis functions per {DG-V} element (No. {DG-V} AO). }
    \label{tab:ch4MP2}
\end{table}

\subsection{Graphene}
\label{sec:Graphene}
\begin{figure}[h!]
    \centering
    \begin{subfigure}{.3\textwidth}
    \centering
    \includegraphics[width=1.\textwidth]{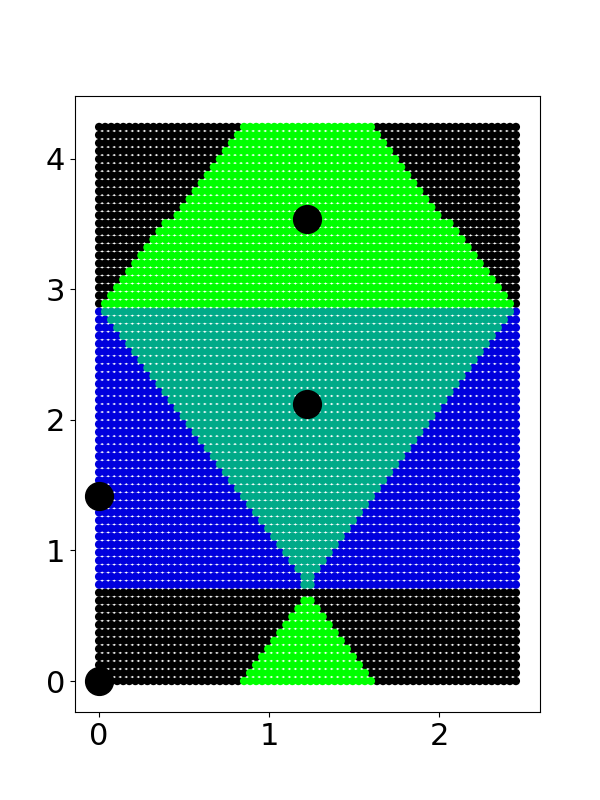}
    \caption{~}
    \label{fig:GrapheneDGV4}
    \end{subfigure}%
    \begin{subfigure}{.3\textwidth}
    \centering
    \includegraphics[width=1.\textwidth]{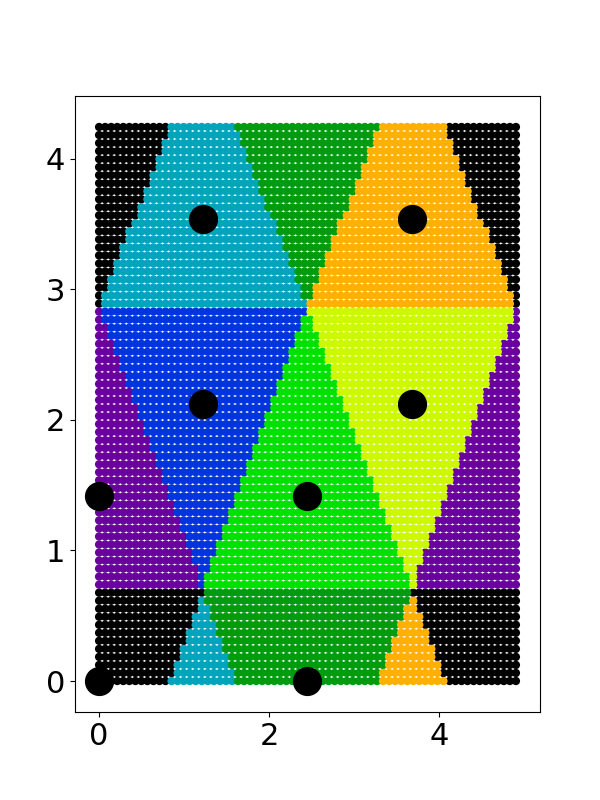}
    \caption{~}
    \label{fig:GrapheneDGV8}
\end{subfigure}
\begin{subfigure}{.3\textwidth}
    \centering
    \includegraphics[width=1.\textwidth]{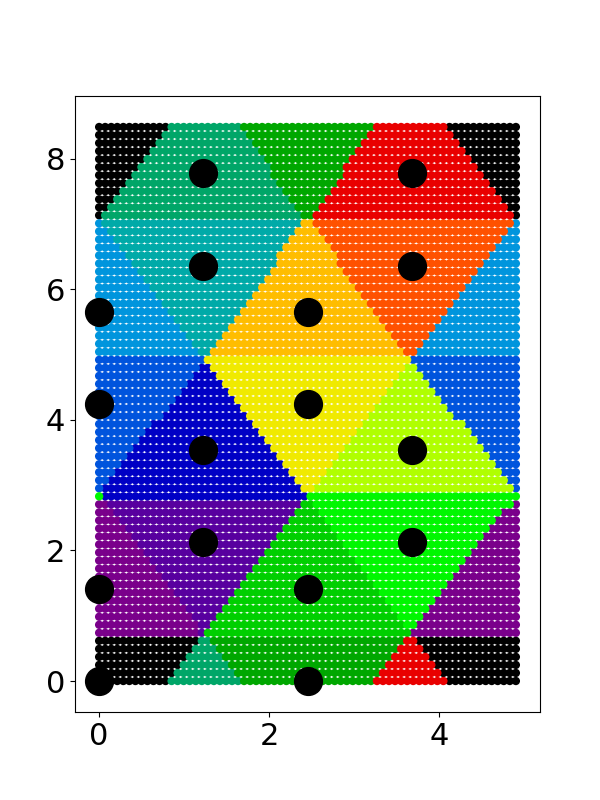}
    \caption{~}
    \label{fig:GrapheneDGV16}
\end{subfigure}
\caption{
{DG-V} partitioning for different orthohombic supercells of graphene:
(a) Minimal orthorhombic supercell four carbon atoms (b) eight carbon atoms (c) sixteen carbon atoms.}
\label{fig:graphene_voronoi}
\end{figure}

In contrast to the single-molecule simulation with a large vacuum package presented in the previous section, the planewave dual basis is most natural for describing  periodic systems (e.g., crystalline solids).
This allows to conveniently extend quantum-simulation methods to condensed
phase systems of interacting electrons.
Subsequently, we exemplify this by applying the {DG-V} approach to graphene (see Figure~\ref{fig:graphene_voronoi}).%---a perfect two-dimensional carbon nanostructure consisting of a mono layer of carbon atoms~\cite{novoselov2004electric}.

% \begin{figure}
%     \centering
%     \includegraphics[width = 0.3\textwidth]{Graphene.pdf}
%     \caption{Graphical representation of an graphene extract and different supercell options. (1) In blue, the primitive cell, (2) in red, the smallest orthorhombic cell and (3) in green a larger orthohombic cell.}
%     \label{fig:Graphene}
% \end{figure}

% \begin{figure}[h!]
%     \centering
%     \begin{subfigure}{.3\textwidth}
%     \centering
%     \includegraphics[width=1.\textwidth]{Images/vdg-small1.png}
%     \caption{~}
%     \label{fig:TiledAtoms}
%     \end{subfigure}%
%     \begin{subfigure}{.3\textwidth}
%     \centering
%     \includegraphics[width=1.\textwidth]{Images/vdg-large1.png}
%     \caption{~}
%     \label{fig:SupercellVoronoi}
% \end{subfigure}
% \caption{(a) Minimal orthorhombic supercell for graphene with four carbon atoms (black atoms). (b) Orthorhombic supercell for graphene with 16 carbon atoms (black atoms).}
% \label{fig:graphene_voronoi}
% \end{figure}

%{\color{red}( xiaojie: Figure 15 and Figure 14 are talking about the same thing. Maybe use Figure 15 only? It has more information.)}

Aligning with the previous investigations, we start with energy computations at the mean-field level of theory. 
We consider graphene with a gth-pade psuedopotential provided in PySCF. 
Using the minimal basis set, gth-szv, we compute graphene for differently sized orthohombic cells containing up 16 atoms. %{\color{red} (Xiaojie: possibly 24 atoms)}.

\begin{figure}[h!]
    \centering
    \begin{subfigure}{.5\textwidth}
    \centering
    \includegraphics[width=1.\textwidth]{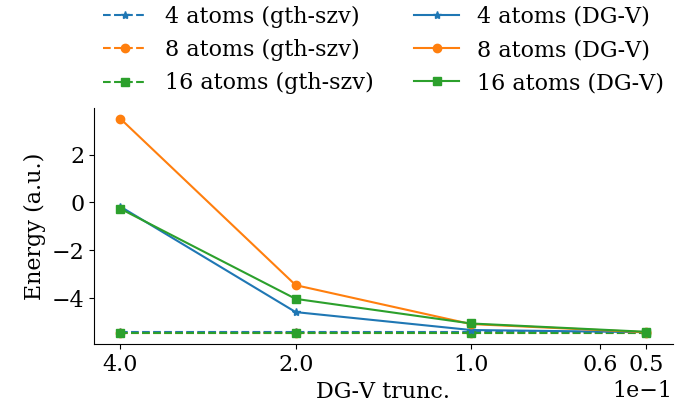}
    \caption{~}
    \label{fig:nb_graphene}
    \end{subfigure}%
    \begin{subfigure}{.5\textwidth}
    \centering
    \includegraphics[width=1.\textwidth]{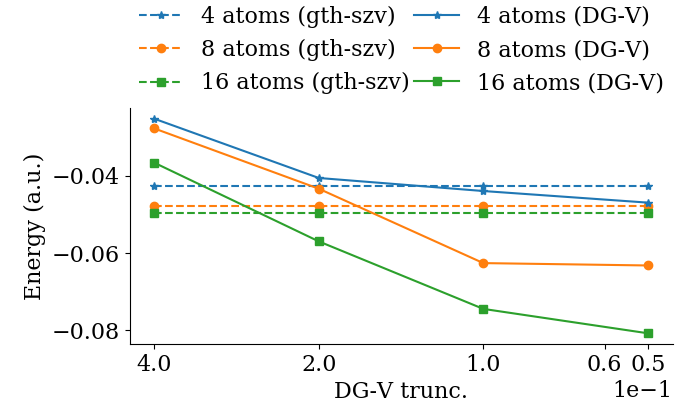}
    \caption{~}
    \label{fig:hf_graphene}
    \end{subfigure}
    \begin{subfigure}{.5\textwidth}
    \centering
    \includegraphics[width=1.\textwidth]{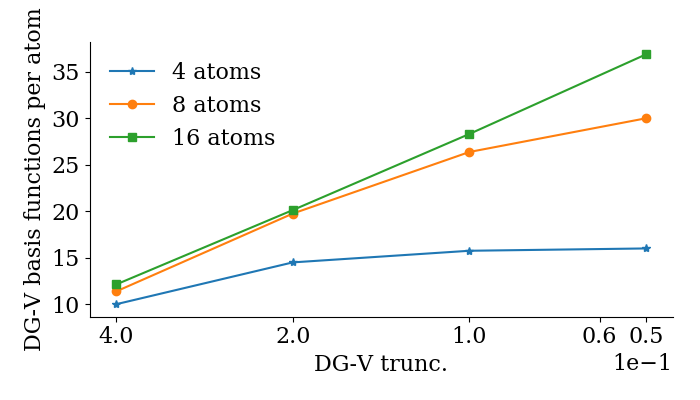}
    \caption{~}
    \label{fig:mp2_graphene}
    \end{subfigure}
\caption{Different quantities as functions of the imposed relative truncation tolerance for the singular values in the construction of the {DG-V} basis functions ({DG-V} trunc): (a) Hartree--Fock energy per atoms (b) MP2 correlation energy per atom (c) Absolute number of {DG-V} basis functions per atom. %{\color{red} Xiaojie: these figures will be updated when the data of the large system is available.}\LL{ consistent figure style; larger font. I do not know what the legend in (b) means} 
}
\label{fig:graphene_result}
\end{figure}

We consider three differently sized orthorhombic cells characterizing graphene: a small-size cell (containing four atoms), a middle-size cell (containing eight atoms), and a large-size cell (containing 16 atoms). For each system, we use a vacuum padding of $5$ angstrom along the z-direction, i.e. the length of the different spuercells along the z-direction is always $10$ angstrom to simulate the vacuum. Each of the boxes is discretized with $60 \times 60 \times 150$ grid points. To reduce the number of {DG-V} basis functions, we use the molecular orbitals, instead of atomic orbitals as active space basis to generate {DG-V} basis. The computational results are provided in Figure~\ref{fig:graphene_result}. 

We decrease the relative tolerance of the singular value truncation
from $0.4$ to $0.05$; note that this corresponds to controlling the number of {DG-V} basis functions through the relative-tolerance criterion outlined in Section~\ref{sec:control}. All calculations are performed using the gth-szv basis set as active space basis in the {DG-V} procedure. We find that the number of {DG-V} basis functions per element will increase as the relative truncation tolerance decreases, see Figure~\ref{fig:graphene_result}(c). Meanwhile, the Hartree--Fock energy of the considered systems obtained by means of the {DG-V} formalism decreases and eventually crosses the Hartree--Fock energy obtain by the periodic boundary condition module in PySCF, see Figure~\ref{fig:graphene_result}(a). The MP2 correlation energy of {DG-V} basis set converges much faster than the mean-field energy. It is surprising that the MP2 correlation energy of the {DG-V} procedure is already close to the result obtained by the active space basis set discretization when the relative truncation tolerance is set to be $0.4$ and the corresponding number of basis functions per element is around $12$. We eventually observe the crossing of the MP2 correlation energies as the relative truncation tolerance becomes smaller than $0.1$, see Figure~\ref{fig:graphene_result}(b). 
%

% We furthermore apply the VdG procedure to selected Pople and correlation consistent basis sets, e.g., ...
% The results are listed in Table~\ref{}.

%Aside from applying VdG to atomic orbitals, it is also possible to choose molecular orbitals as active space basis set. 
%
%This should dramatically reduce the number of VdG basis functions per element (If we cannot confirm then we drop this)

\section{Conclusion}

In this article we have introduced the discontinuous Galerkin formalism with Voronoi partitioning ({DG-V}), an extension to the discontinuous Galerkin procedure with rectangular partitioning ({DG-R}) presented in Ref.~\cite{mcclean2020discontinuous}. 
The aim of both {DG-V} and {DG-R} is to provide a smooth interpolation between the regime of a primitive basis sets (such as a planewave dual basis set) and an activate space basis set (such as a Gaussian basis set). 
The Hamiltonian, and in particular the ERI tensor in the {DG-V} basis set takes a block-diagonal form, which leads to asymptotically efficient quantum algorithms, but the preconstant can be significantly reduced compared to that of the primitive basis set. 
Compared to {DG-R}, the {DG-V} method  provides a more flexible partitioning procedure allowing to apply it to molecular and solid state systems of arbitrary geometry. 
{DG-V} always produces an orthonormal basis set, and the block structure of the basis set naturally leads to easy and efficient implementations of density fitting techniques, which we will explore in the future. 

Although the {DG-V} procedure dramatically reduces the number of basis functions compared the primitive basis set, when used for classical quantum chemistry calculations at the correlated level (e.g. MP2 or CCSD calculations), we find that the increase of the total number of basis functions compared to the standard active orbital implementation is still significant, at least for systems of relatively small sizes. This is particularly the case since our current implementation in PySCF cannot efficiently take into account the sparsity pattern of the ERI tensor in the {DG-V} basis for calculations at the correlated level. We expect that in the near future, the {DG-V} scheme could be more useful when effectively taking advantage of such sparsity pattern, such in the settings of quantum embedding methods~\cite{KniziaChan2012,SunChan2016}.

\section*{Acknowledgments:} 

This work was partially supported by the Department of Energy under Grant No. DE-SC0017867, the National Science Foundation under the Quantum Leap Challenge Institutes, the Google Quantum Research Award (L.L.), and by the Air Force Office of Scientific Research under award number FA9550-18-1-0095 (F.F.,X.W.,L.L.).

%\newpage
%\section*{References}
\newcommand{\newblock}{}
\bibliographystyle{abbrv}
\bibliography{lib}
%\bibliography{../../../Dropbox/Bibliography/reference,lib}

\newpage

\include{supp}

\end{document}

%% file: supp.tex
\appendix

\section{Supercell model and the planewave dual basis set}\label{app:planewavedual}

In this section we briefly describe the use of the planewave dual basis set for quantum many-body calculations. Associated with the Bravais lattice is a reciprocal lattice 
\[
\mathbb{L}^* = \left\{
\vG = \left( \frac{2\pi}{L_1} i_1,\frac{2\pi}{L_2} i_2,\frac{2\pi}{L_3} i_3 \right),
\quad (i_1,i_2,i_3)\in \ZZ^3
\right\}.
\]
It is convenient to introduce two sets of  grids, one in the Fourier space and one in the real space. The one in the Fourier space is indexed by a subset of $\mathbb{L}^*$, denoted by
\[
\GG = \left\{
\vG = \left( \frac{2\pi}{L_1} i_1,\frac{2\pi}{L_2} i_2,\frac{2\pi}{L_3} i_3 \right),
\quad
-\frac{n_1}{2}\le i_1<\frac{n_1}{2},
-\frac{n_2}{2}\le i_2<\frac{n_2}{2},
-\frac{n_3}{2}\le i_3<\frac{n_3}{2}
\right\},
\]
where $n_1$, $n_2$, and $n_3$ are assumed to be even for simplicity. The second is a Cartesian grid in the real space is denoted by
\[
\XX = \left\{
\vr=\left(\frac{i_1}{n_1} L_1, \frac{i_2}{n_2} L_2, \frac{i_3}{n_3}
L_3\right):
0\le i_1<n_1, 0\le i_2<n_2, 0\le i_3<n_3
\right\}.
\]
The total number of grid points of both grids is equal to 
\begin{equation}
N_g = n_1 n_2 n_3.
\label{eqn:number_grid}
\end{equation}
For simplicity we shall give each point in $\XX$ a lexicographic order so that $\XX=\{\vr_\mu\}_{\mu=1}^{N_g}$.

% Note that in DFT calculations, the Fourier grid is often confined within a sphere characterized by a single number $E_{\text{cut}}$. So the calculation restricts to only the Fourier modes indexed by
% \[
% \GG_{\text{cut}} := \left\{\vG\in\mathbb{L}^*: \frac{1}{2} |\vG|^2 \le E_{\text{cut}} \right\}\subset \mathbb{L}^*.
% \]
% For simplicity we do not consider this spherical grid here, and assume $\GG_{\text{cut}}=\GG$.

Consider a smooth single-particle function $\psi(\vr)$, its value at any point $\vr\in\XX$ is given by
\begin{equation}
  \psi(\vr) =\frac{1}{|\Omega|} \sum_{\vG\in\GG}
  \hat{\psi}(\vG)  \exp(\I \vG \cdot \vr).
  \label{eqn:psi}
\end{equation}
The Fourier coefficients $\{ \hat{\psi}(\vG), \vG\in\GG \}$ can be computed using the samples of $\psi(\vr)$
at $\XX$:
%%\llin{consistency in terms of  convention of complex conjugation}
\begin{equation}
  \hat{\psi}(\vG) 
  = \int_\Omega  \exp(-\I \vG \cdot  \vr) \psi(\vr) \ud \vr
  = \dOmega \sum_{\vr\in\XX} \psi(\vr) \exp(-\I \vG \cdot \vr).
  \label{eqn:hatpsi}
\end{equation}
Here  $\dOmega:=|\Omega|/N_g$ the size of the volume element.

We now define a set of functions in the real space as
\begin{equation}
  \chi_{\vr'}(\vr) =
  \frac{1}{{\sqrt{N_g|\Omega|}}}\sum_{\vG\in\GG} \exp(\I \vG\cdot(\vr-\vr')).
  \label{eqn:fourierdual}
\end{equation}
In particular, they can be viewed as the numerical $\delta$-functions on the discrete set
$\XX$, satisfying
\begin{equation}
  \chi_{\vr'}(\vr) = \dOmega^{-\frac12} \delta_{\vr,\vr'},
  \quad \vr, \vr'\in \XX.
  \label{eqn:fourierdualgrid}
\end{equation}
We introduce a simplified notation $\chi_{\mu}(\vr):=\chi_{\vr_{\mu}}(\vr),\vr_{\mu}\in \XX, 1\le\mu\le N_g$. Then we may verify the orthonormality condition \[
\begin{split}
  \int_{\Omega} \overline{\chi}_{\mu}(\vr) \chi_{\nu}(\vr) \ud \vr =&\frac{1}{N_g \abs{\Omega}} \int_{\Omega} \sum_{\vG,\vG'\in\GG} \exp(-\I \vG\cdot(\vr-\vr_\mu)) \exp(\I \vG'\cdot(\vr-\vr_\nu))  \ud \vr\\
=&\sum_{\vG\in\GG} \frac{1}{N_g} \exp(\I \vG\cdot(\vr_{\mu}-\vr_\nu))=\delta_{\mu,\nu}.
\end{split}
\]
Furthermore, the aforementioned $\psi(\vr)$ can be expanded using its nodal values as
\begin{equation}
  \psi(\vr) =\dOmega^{\frac12} \sum_{\vr_{\mu}\in \XX} \chi_{\mu}(\vr) \psi_{\mu},
  \label{eqn:collocation}
\end{equation}
with the short hand notation $\psi_{\mu}:=\psi(\vr_{\mu})$.   The basis set $\{\chi_{\mu}(\vr)\}_{\mu=1}^{N_g}$ is called the planewave dual basis set.

% In quantum chemistry, the basis set $\{\chi_{\mu}(\vr)\}_{\mu=1}^{N_g}$ goes under the name of pseudospectral (PS) method~\cite{}, discrete variable representation (DVR)~\cite{}, periodic sinc ($\operatorname{psinc}$) basis set \cite{SkylarisHaynesMostofiEtAl2005},  the planewave dual basis set \cite{MardirossianMcClainChan2018,babbush2018low}. 
For a given function, the nodal representation~\eqref{eqn:collocation} allows one to identify its function values evaluated at Cartesian grid points $\XX$ with the expansion coefficients under the planewave dual basis set. This is particularly convenient when computing integrals. 
% 
% Using the planewave dual basis set, a quantum many-body wavefunction in the first quantization formulation can be written as
% \[
% \Psi(\vr_1,\ldots,\vr_N)=\dOmega^{\frac{N}{2}}\sum_{\mu_1,\ldots,\mu_N=1}^{N_g} \Psi_{\mu_1,\ldots,\mu_N} \prod_{i=1}^N \chi_{\mu_i}(\vr_i),
% \]
% where $\Psi_{\vr'_1,\ldots,\vr'_N}$ is an anti-symmetric tensor of size $(N_g)^N$. We shall omit the range of summation when the context is clear. The normalization condition requires
% \[
% \int |\Psi(\vr_1,\ldots,\vr_N)|^2 \ud \vr_1\cdots\vr_N=\dOmega^{N} \sum_{\mu_1,\ldots,\mu_N}  \abs{\Psi_{\mu_1,\ldots,\mu_N}}^2=1.
% \]
% 
% In the special case of a single Slater determinant, we first define a set of single-particle orbitals 
% \[
% \psi_{p}(\vr)=\dOmega^{\frac12} \sum_{\mu} \psi_{\mu,p} \chi_{{\mu}}(\vr),
% \]
% satisfying the orthogonality condition
% \[
% \int_{\Omega} \overline{\psi}_{p}(\vr)\psi_{q}(\vr)\ud \vr=\dOmega\sum_{\mu} \overline{\psi}_{\mu,p} \psi_{\mu,q} = \delta_{p,q}.
% \]
% This defines a coefficient matrix as Slater determinant 
% \[
% \Psi_{\mu_1,\ldots,\mu_N}=\frac{1}{\sqrt{N!}}\begin{vmatrix}
% \psi_{\mu_1,1} & \cdots & \psi_{\mu_N,1}\\
% \vdots & \ddots & \vdots \\
% \psi_{\mu_1,N} & \cdots & \psi_{\mu_N,N}\\
% \end{vmatrix}.
% \]

The supercell model also requires us to replace the Coulomb kernel in the free space by the Ewald potential \cite{FraserFoulkesRajagopalEtAl1996}, denoted by $v^{\per}(\vr-\vr')$, defined in terms of its Fourier coefficients as
\[
v^{\per}(\vr-\vr')=\left(\frac{4\pi}{\abs{\Omega}} \sum_{\vG\ne \mathbf{0},\vG\in\GG} \frac{1}{\abs{\vG}^2}\exp\left(\I \vG \cdot (\vr-\vr')\right)\right)-\xi=:\sum_{\vG\in\GG} \hat{v}^{\per}(\vG) \exp\left(\I \vG \cdot (\vr-\vr')\right).
\]
Here
\begin{equation}
\hat{v}^{\per}(\vG)=\begin{cases}
\frac{1}{\abs{\Omega}}\frac{4\pi}{\abs{\vG}^2}, & \vG\ne \mathbf{0},\\
-\xi,& \vG= \mathbf{0},
\end{cases}
\end{equation}
and $\xi$ is a constant. For a cubic cell of length $L$, we have $\xi=-\nu_M/L$ where $\nu_M$ is the the Madelung constant of the supercell  \cite{FraserFoulkesRajagopalEtAl1996}. For simplicity we have absorbed the shift due to the Madelung constant into the definition of the Ewald potential, which can significantly reduce the finite size error of mean-field energies and orbital energies. 

% The Madelung constant converges slowly as $\Or(L^{-1})$ with respect to the length of the supercell (assuming a simple cubic cell with $L_1=L_2=L_3=L$). Note that it is also common to define the Ewald potential such that $\hat{v}^{\per}(\boldsymbol 0)=0$. In such a case, a post-processing step involving the Madelung constant is often necessary to accelerate the convergence towards the free space limit (for molecules) and thermodynamic limit (for solids) \cite{MakovPayne1995,FraserFoulkesRajagopalEtAl1996}. This is particularly important for the evaluation of the Fock exchange energy \cite{GygiBaldereschi1986,SundararamanArias2013,McClainSunChanEtAl2017}.

Note that both the planewave basis and the planewave dual basis are delocalized functions in the supercell. Hence a Galerkin formulation generally leads to dense matrices. For instance, the matrix due to electron-nuclei interaction\[
V^{\per}_{\text{en},\mu\nu}=-\sum_{I}Z_I\int_{\Omega} \overline{\chi}_{{\mu}}(\vr) v^{\per}(\vr-\vR_I) \chi_{{\nu}}(\vr)\ud  \vr
\]
is a dense matrix of size $N_g\times N_g$. However, the sparsity can be improved using the pseudospectral formulation by replacing an integral by a quadrature using the nodal values of functions for $\vr\in\XX$, i.e.
\[
V^{\per}_{\text{en},\mu\nu} \approx
-\dOmega \sum_{I}Z_I\sum_{\vr\in\XX} \overline{\chi}_{{\mu}}(\vr) v^{\per}(\vr-\vR_I) \chi_{{\nu}}(\vr):= \left(-\sum_{I}Z_I v^{\per}(\vr_{\mu}-\vR_I) \right)\delta_{\mu\nu},
\]
which is a diagonal matrix. 
Here we have used \cref{eqn:fourierdualgrid}. Similarly, the two-electron repulsion integral (ERI) matrix elements can be written as  \begin{equation}
  \braket{\mu \nu|\lambda\gamma}=\int_{\Omega} \int_{\Omega}
  \overline{\chi}_{\mu}(\vr)\overline{\chi}_{\nu}(\vr')  v^{\per}(\vr-\vr')   \chi_{\lambda}(\vr)\chi_{\gamma}(\vr') \ud \vr \ud \vr' \approx v^{\per}(\vr_{\mu}-\vr_{\nu}) \delta_{\mu,\lambda}\delta_{\nu,\gamma}.
\label{eqn:ERI_diagonal}
\end{equation}
 If we reshape the ERI tensor as a matrix by considering the multi-indices $(\mu\nu)$ and $(\lambda\gamma)$, respectively, the resulting matrix is a diagonal matrix. Hence \cref{eqn:ERI_diagonal} will be referred to as the diagonal representation of ERIs. This gives the Hamiltonian in the planewave dual basis in \cref{eqn:manybody_pwdual} in the second quantized form. Note that to reduce the number of planewave dual basis sets, the electron-nuclei interaction can also be replaced by a pseudopotential formulation. In this work, we employ the GTH pseudopotential~\cite{GoedeckerTeterHutter1996}  as implemented in PySCF.

\section{H$_4$ transition from square to linear geometry}
In the following section we will present further numerical results that align with the results presented in Section~\ref{sec:P4toD4} but extend to further basis sets.
%\LL{ present the figures in the order that appears in the text.} 

\subsection{Condition number, grid spacing and egg-box effect}

\begin{figure}[h!]
    \centering
    \begin{subfigure}{.5\textwidth}
    \centering
    \includegraphics[width=1.\textwidth]{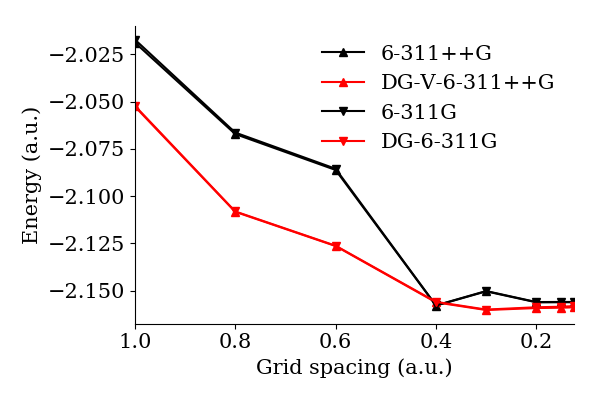}
    \caption{~}
    \end{subfigure}%
    \begin{subfigure}{.5\textwidth}
    \centering
    \includegraphics[width=1.\textwidth]{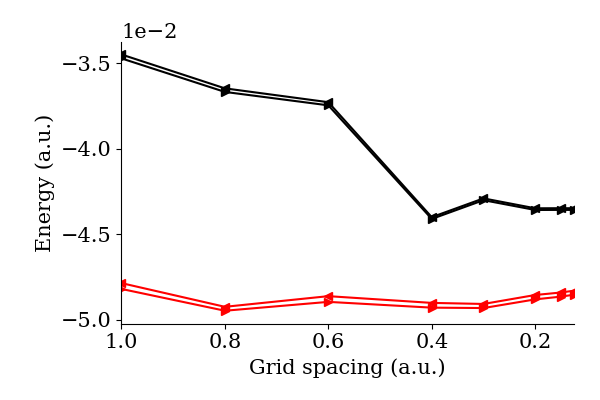}
    \caption{~}
\end{subfigure}
\caption{
Convergence of mean field (a) and MP2 correlation energy (b) as functions of the grid spacing for the Pople basis sets 6-311G and 6-311++G. }
\label{fig:Grid_conv_6311g}
\end{figure}

In this subsection we present additional data on the (near computational) linear dependence, the convergence of the Hartree--Fock energy as a function of the grid spacing and the ``egg-box'' effect. As remarked in Section~\ref{sec:P4toD4}, we have investigated the convergence behavior of the Hartree--Fock and MP2 energies as functions of the grid spacing. We found that at $0.2$ (a.u.), the Hartree--Fock and MP2 energies are sufficiently converged, see Figure~\ref{fig:Grid_conv_6311g}.

% \begin{figure}[h!]
%     \centering
%     \includegraphics[width=\textwidth]{Images/Grid_conv_6311g.png}
%     \caption{Convergence of mean field (left) and MP2 correlation energy (right) as functions of the grid spacing for the Pople basis sets 6-311G and 6-311++G. }
%     \label{fig:Grid_conv_6311g}
% \end{figure}

%
Next we show an additional computation of the ``egg box'' effect for the DG-V procedure using the cc-pVDZ basis set. We confirm that the egg-box effect caused by the DG-V procedure is on the order of $3\times 10^{-5}$ Hartree and is negligible. The size of the error is also comparable to that in the pbc (periodic boundary condition) module of PySCF.

\begin{figure}[h!]
    \centering
    \begin{subfigure}{.33\textwidth}
    \centering
    \includegraphics[width=1.\textwidth]{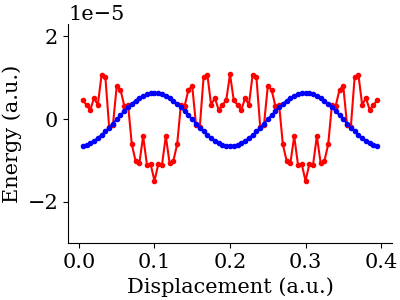}
    \caption{~}
    \end{subfigure}%
    \begin{subfigure}{.33\textwidth}
    \centering
    \includegraphics[width=1.\textwidth]{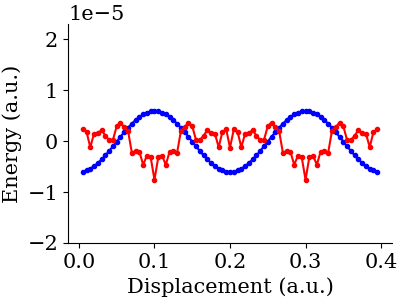}
    \caption{~}
\end{subfigure}
\begin{subfigure}{.33\textwidth}
    \centering
    \includegraphics[width=1.\textwidth]{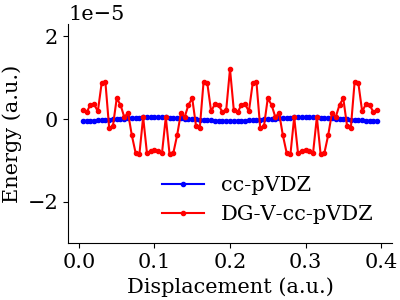}
    \caption{~}
\end{subfigure}
\caption{
``Egg-box" effect for H$_4$ with $\alpha = 60$ degree with cc-pVDZ as active basis set and an underlying real-space grid with grid spacing of $0.2$ (a.u.). Different energy contributions (a) total energy (b) mean field energy (c) MP2 correlation energy are plotted as function of the displacement along the horizontal symmetry axis (cf Figure~\ref{fig:P4toD4}). }
\label{fig:EggBoxccpvdz}
\end{figure}

% \begin{figure}[h!]
%     \centering
%     \includegraphics[width=\textwidth]{Images/EggBox_ccpvdz_fine1.png}
%     \caption{``Egg-box" effect for H$_4$ with $\alpha = 60$ degree with cc-pVDZ as active basis set and an underlying real-space grid with grid spacing of $0.2$ (a.u.). Different energy contributions are plotted as function of the displacement along the horizontal symmetry axis (cf Figure~\ref{fig:P4toD4}).}
%     \label{fig:EggBoxccpvdz}
% \end{figure}

\begin{figure}[h!]
    \centering
    \begin{subfigure}{.49\textwidth}
    \centering
    \includegraphics[width=1.\textwidth]{Images/Cond_No_ccpvdz.png}
    \caption{~}
    \end{subfigure}%
    \begin{subfigure}{.49\textwidth}
    \centering
    \includegraphics[width=1.\textwidth]{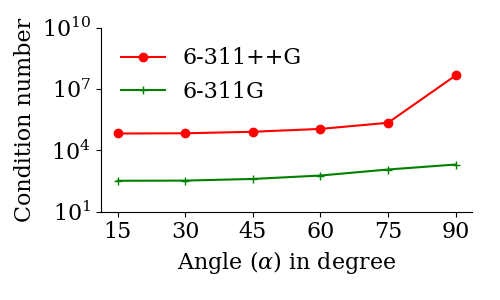}
    \caption{~}
\end{subfigure}
\begin{subfigure}{.49\textwidth}
    \centering
    \includegraphics[width=1.\textwidth]{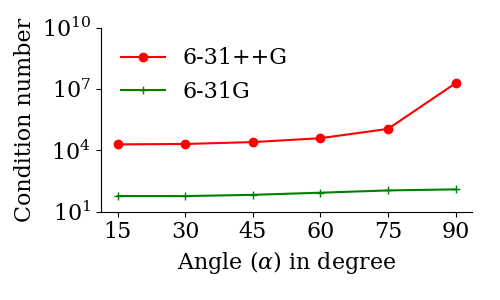}
    \caption{~}
\end{subfigure}
\begin{subfigure}{.49\textwidth}
    \centering
    \includegraphics[width=1.\textwidth]{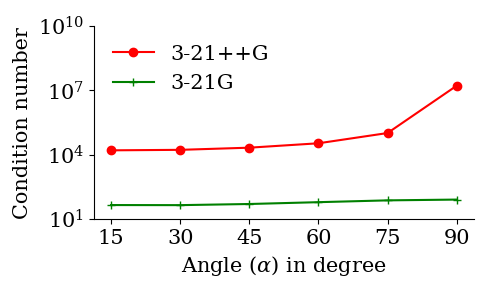}
    \caption{~}
\end{subfigure}
\caption{
Condition number of the overlap matrix as a function of the parameter $\alpha$ for different choices of basis functions for the H$_4$ molecule.}
\label{fig:ConditionNumberDiffBasis}
\end{figure}

\pagebreak
We have furthermore the computed condition number of the overlap matrix as a function of the system parameter $\alpha$ for different basis sets in Figure~\ref{fig:ConditionNumberDiffBasis}. We see that the adding diffuse basis functions increase in general the condition number, and that the condition number in all cases is significantly higher than one, which is the condition number of DG-V overlap matrix.

% \begin{figure}[h!]
%     \centering
%     \includegraphics[width=0.8\textwidth]{Images/H4_trans_ConNo.png}
%     \caption{Condition number of the overlap matrix as a function of the parameter $\alpha$ for different choices of basis functions for the H$_4$ molecule.}
%     \label{fig:ConditionNumberDiffBasis}
% \end{figure}

% \begin{figure}
%     \centering
%     \includegraphics[width=\textwidth]{Images/mf_P4-to-D4_6311.png}
%     \caption{Caption}
%     \label{fig:PES_mf_6311}
% \end{figure}

% \begin{figure}
%     \centering
%     \includegraphics[width=\textwidth]{Images/mf_P4-to-D4_631.png}
%     \caption{Caption}
%     \label{fig:PES_mf_631}
% \end{figure}

% \begin{figure}
%     \centering
%     \includegraphics[width=\textwidth]{Images/mf_P4-to-D4_321.png}
%     \caption{Caption}
%     \label{fig:PES_mf_321}
% \end{figure}

\subsection{Mean field computations}
\label{app:TruncationSensitivity}

\begin{figure}[h!]
    \centering
    \begin{subfigure}{.49\textwidth}
    \centering
    \includegraphics[width=1.\textwidth]{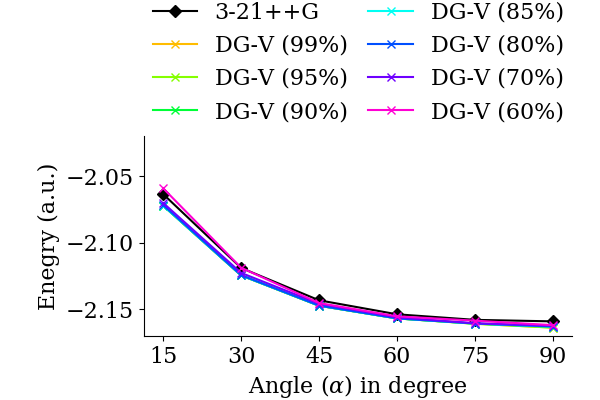}
    \caption{~}
    \end{subfigure}%
    \begin{subfigure}{.49\textwidth}
    \centering
    \includegraphics[width=1.\textwidth]{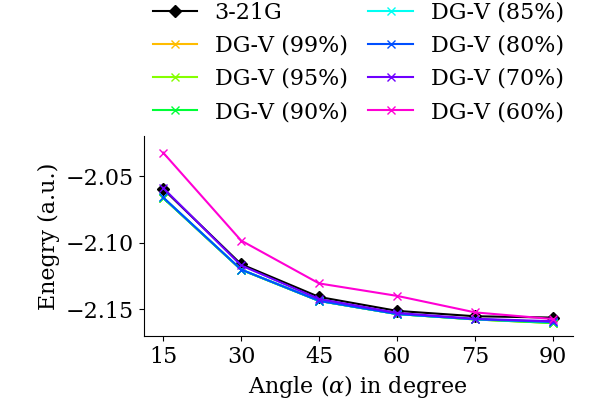}
    \caption{~}
\end{subfigure}
\begin{subfigure}{.49\textwidth}
    \centering
    \includegraphics[width=1.\textwidth]{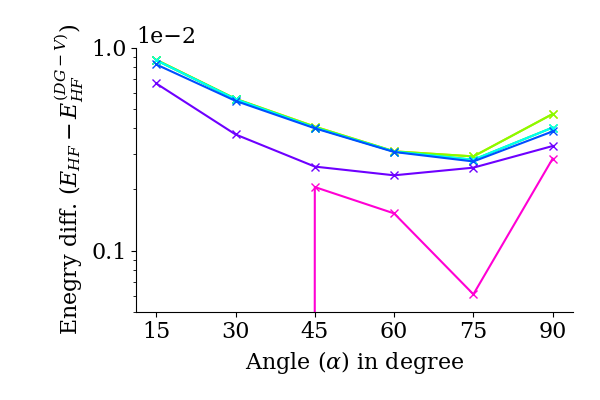}
    \caption{~}
\end{subfigure}
\begin{subfigure}{.49\textwidth}
    \centering
    \includegraphics[width=1.\textwidth]{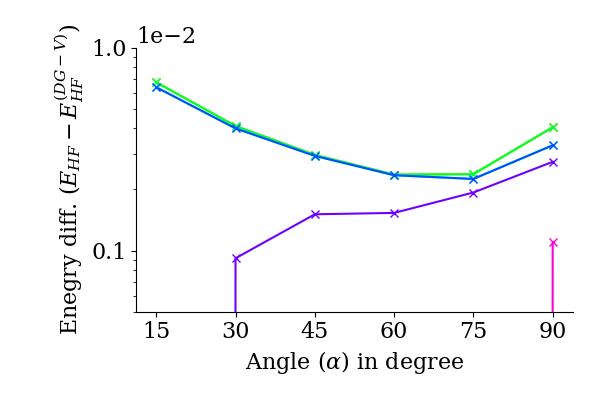}
    \caption{~}
\end{subfigure}
\caption{
Subplots (a) and (b) show the absolute energies for 3-21G and 3-21++G for different truncations of the number of DG-V basis functions. The relative numbers of DG-V basis functions kept per element are reported in percentage. The DG-V solutions are compared to the solution provided by the periodic boundary condition module in PySCF ( depicted by the black dimond markers). The subplots (c) and (d) show the differences of the DG-V solutions to the PySCF solution, i.e. $ E_{\rm HF} - E_{\rm HF}^{\rm (DG-V)}$ on a semilogarithmic scale. Note that as the number of the DG-V basis increases, the energy from the 3-21G and 3-21++G basis set is higher than that from the DG-V basis set.}
\label{fig:MFtruncationApp321}
\end{figure}

\begin{figure}[h!]
    \centering
    \begin{subfigure}{.49\textwidth}
    \centering
    \includegraphics[width=1.\textwidth]{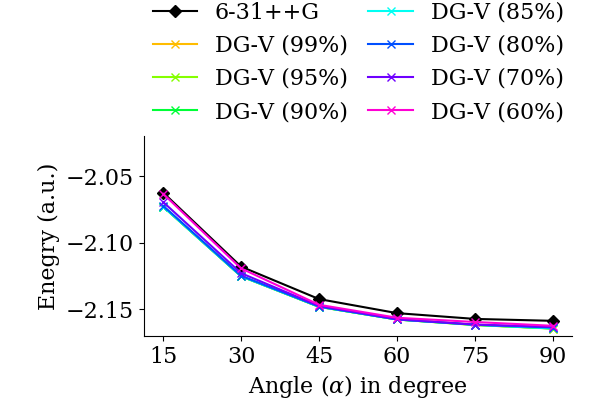}
    \caption{~}
    \end{subfigure}%
    \begin{subfigure}{.49\textwidth}
    \centering
    \includegraphics[width=1.\textwidth]{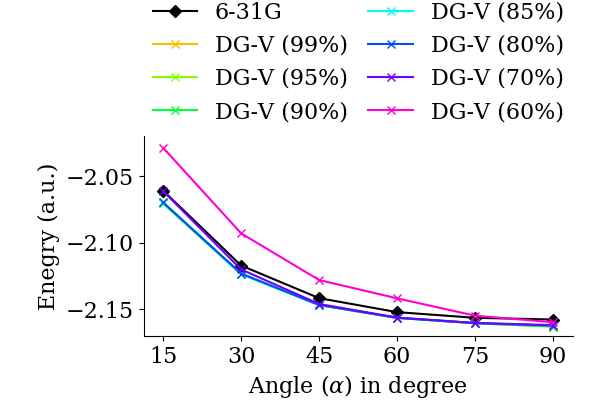}
    \caption{~}
\end{subfigure}
\begin{subfigure}{.49\textwidth}
    \centering
    \includegraphics[width=1.\textwidth]{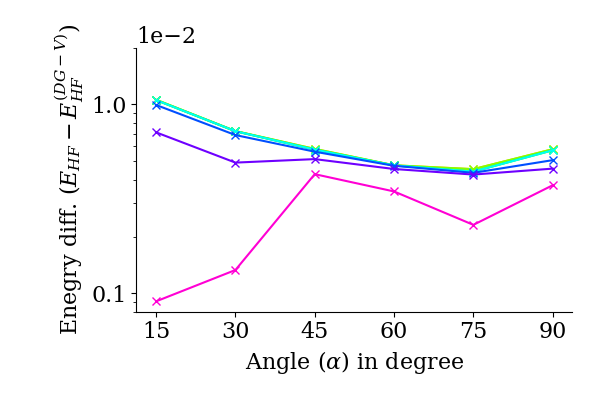}
    \caption{~}
\end{subfigure}
\begin{subfigure}{.49\textwidth}
    \centering
    \includegraphics[width=1.\textwidth]{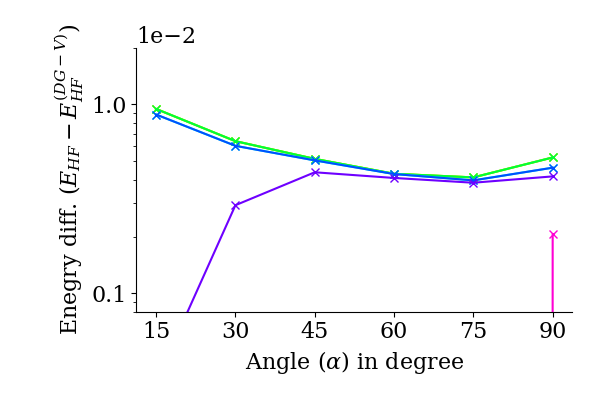}
    \caption{~}
\end{subfigure}
\caption{
Subplots (a) and (b) show the absolute energies for 6-31G and 6-31++G for different truncations of the number of DG-V basis functions. The relative numbers of DG-V basis functions kept per element are reported in percentage. The DG-V solutions are compared to the solution provided by the periodic boundary condition module in PySCF ( depicted by the black dimond markers). The subplots (c) and (d) show the differences of the DG-V solutions to the PySCF solution, i.e. $ E_{\rm HF} - E_{\rm HF}^{\rm (DG-V)}$ on a semilogarithmic scale. Note that as the number of the DG-V basis increases, the energy from the 6-31G and 6-31++G basis set is higher than that from the DG-V basis set.}
\label{fig:MFtruncationApp631}
\end{figure}

\begin{figure}[h!]
    \centering
    \begin{subfigure}{.49\textwidth}
    \centering
    \includegraphics[width=1.\textwidth]{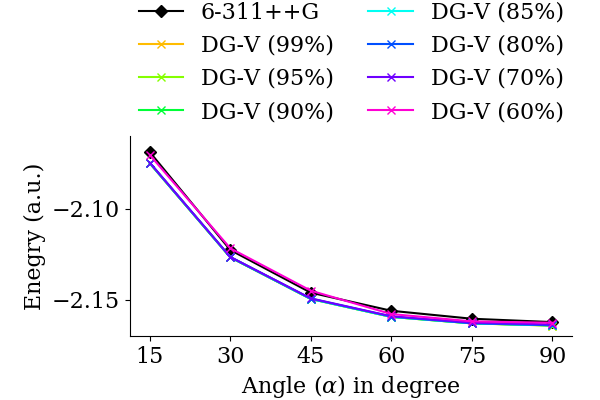}
    \caption{~}
    \end{subfigure}%
    \begin{subfigure}{.49\textwidth}
    \centering
    \includegraphics[width=1.\textwidth]{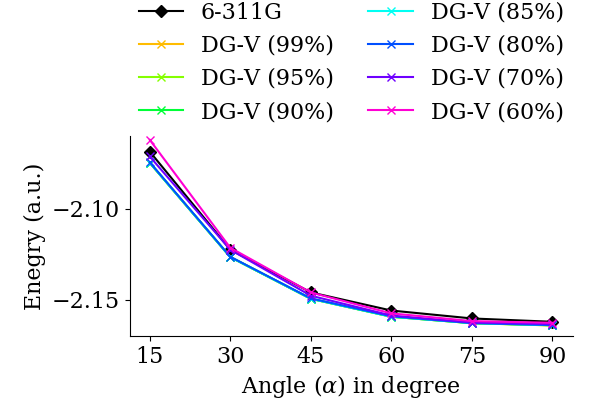}
    \caption{~}
\end{subfigure}
\begin{subfigure}{.49\textwidth}
    \centering
    \includegraphics[width=1.\textwidth]{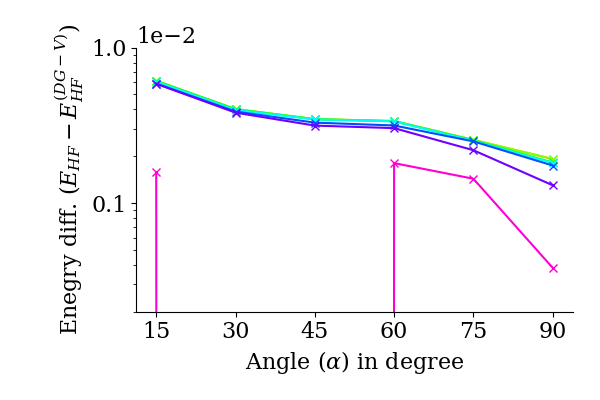}
    \caption{~}
\end{subfigure}
\begin{subfigure}{.49\textwidth}
    \centering
    \includegraphics[width=1.\textwidth]{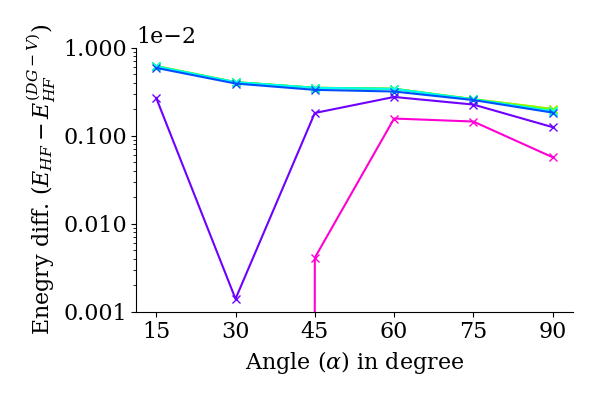}
    \caption{~}
\end{subfigure}
\caption{
Subplots (a) and (b) show the absolute energies for 6-311G and 6-311++G for different truncations of the number of DG-V basis functions. The relative numbers of DG-V basis functions kept per element are reported in percentage. The DG-V solutions are compared to the solution provided by the periodic boundary condition module in PySCF ( depicted by the black dimond markers). The subplots (c) and (d) show the differences of the DG-V solutions to the PySCF solution, i.e. $ E_{\rm HF} - E_{\rm HF}^{\rm (DG-V)}$ on a semilogarithmic scale. Note that as the number of the DG-V basis increases, the energy from the 6-311G and 6-311++G basis set is higher than that from the DG-V basis set.}
\label{fig:MFtruncationApp6311}
\end{figure}

\begin{figure}[h!]
    \centering
    \begin{subfigure}{.49\textwidth}
    \centering
    \includegraphics[width=1.\textwidth]{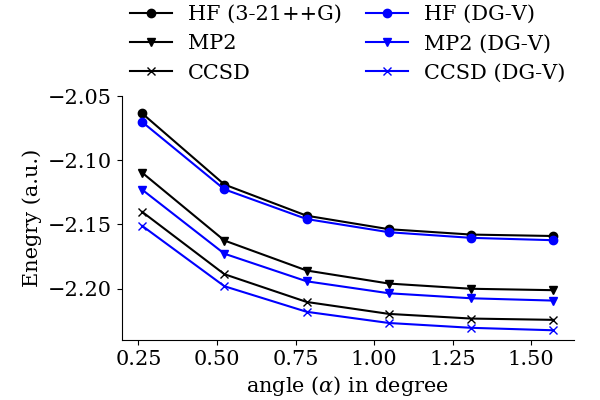}
    \caption{~}
    \end{subfigure}%
    \begin{subfigure}{.49\textwidth}
    \centering
    \includegraphics[width=1.\textwidth]{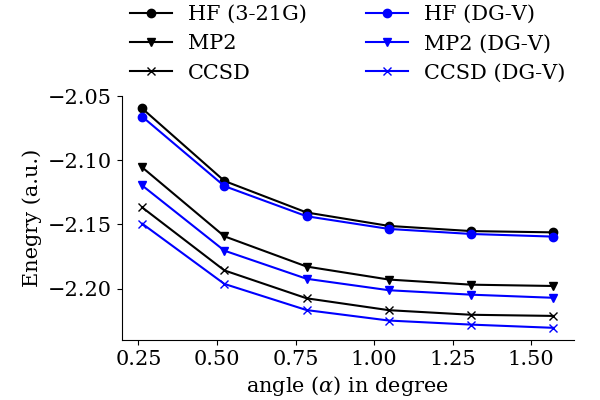}
    \caption{~}
\end{subfigure}
\begin{subfigure}{.49\textwidth}
    \centering
    \includegraphics[width=1.\textwidth]{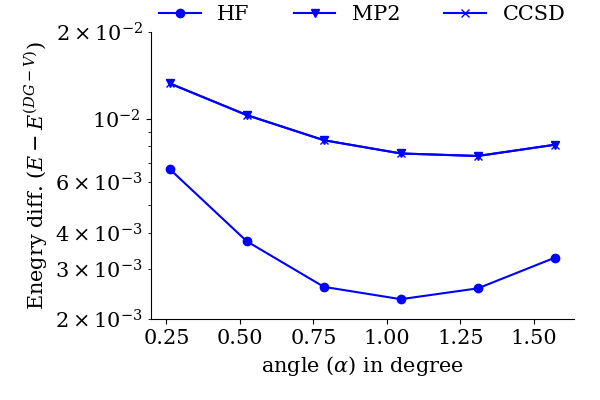}
    \caption{~}
\end{subfigure}
\begin{subfigure}{.49\textwidth}
    \centering
    \includegraphics[width=1.\textwidth]{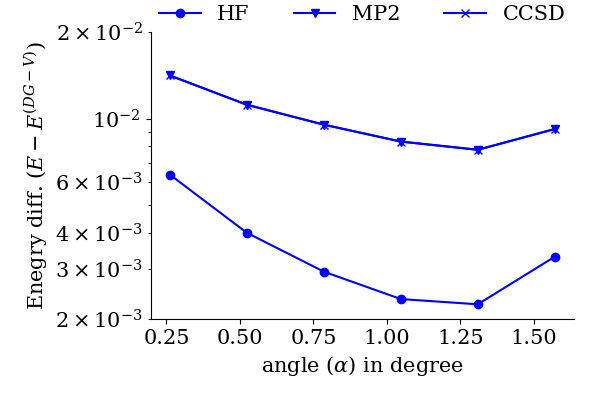}
    \caption{~}
\end{subfigure}
\caption{
Subplots (a) and (b) show the absolute energies for mean field, MP2 and CCSD energy computations. The truncations of the number of DG-V basis functions are extracted from previous mean field computations and was \emph{a priori} set to be 70\% and 80\% for 3-21++G and 3-21G, respectively. The DG-V solutions are compared to the solution provided by the periodic boundary condition module in PySCF ( depicted by the black curves).
The subplots (c) and (d) show the differences of the DG-V solutions to the respective PySCF solution, i.e. $ E_{\rm HF} - E_{\rm HF}^{\rm (DG-V)}$ on a semilogarithmic scale. }
\label{fig:correlation321G}
\end{figure}

\begin{figure}[h!]
    \centering
    \begin{subfigure}{.49\textwidth}
    \centering
    \includegraphics[width=1.\textwidth]{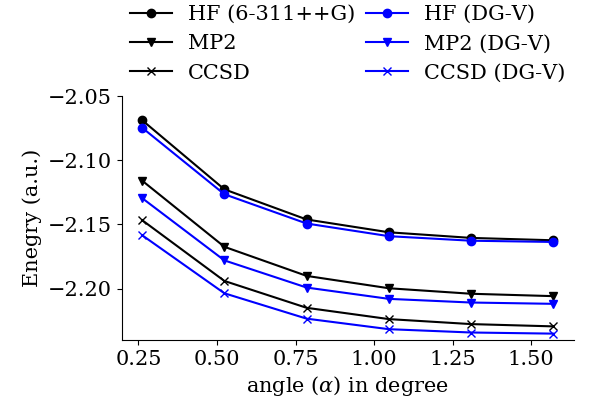}
    \caption{~}
    \end{subfigure}%
    \begin{subfigure}{.49\textwidth}
    \centering
    \includegraphics[width=1.\textwidth]{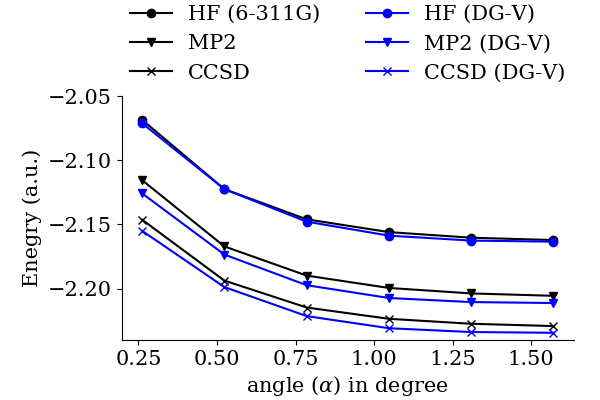}
    \caption{~}
\end{subfigure}
\begin{subfigure}{.49\textwidth}
    \centering
    \includegraphics[width=1.\textwidth]{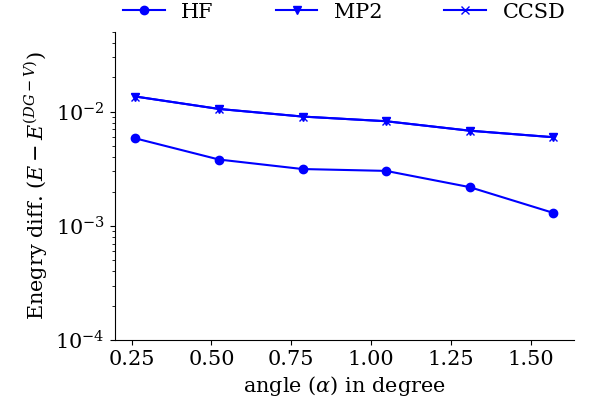}
    \caption{~}
\end{subfigure}
\begin{subfigure}{.49\textwidth}
    \centering
    \includegraphics[width=1.\textwidth]{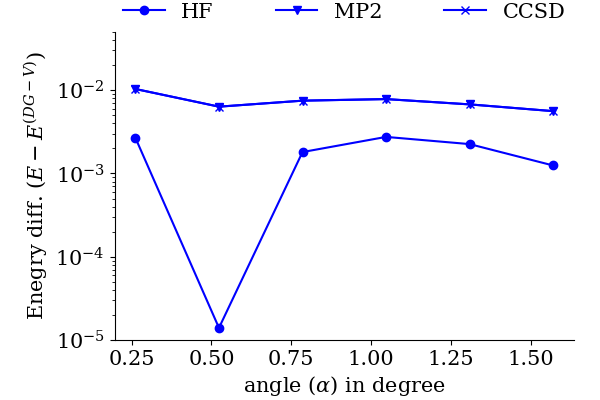}
    \caption{~}
\end{subfigure}
\caption{
Subplots (a) and (b) show the absolute energies for mean field, MP2 and CCSD energy computations. The truncations of the number of DG-V basis functions are extracted from previous mean field computations and was \emph{a priori} set to be 70\% and 70\% for 6-311++G and 6-311G, respectively. The DG-V solutions are compared to the solution provided by the periodic boundary condition module in PySCF ( depicted by the black curves).
The subplots (c) and (d) show the differences of the DG-V solutions to the respective PySCF solution, i.e. $ E_{\rm HF} - E_{\rm HF}^{\rm (DG-V)}$ on a semilogarithmic scale. }
\label{fig:correlation6311G}
\end{figure}

\begin{figure}[h!]
    \centering
    \begin{subfigure}{.49\textwidth}
    \centering
    \includegraphics[width=1.\textwidth]{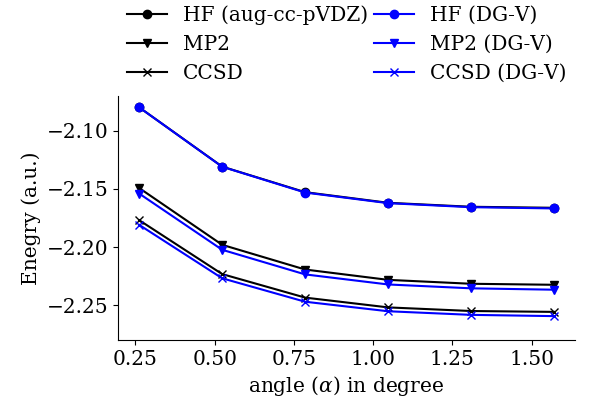}
    \caption{~}
    \end{subfigure}%
    \begin{subfigure}{.49\textwidth}
    \centering
    \includegraphics[width=1.\textwidth]{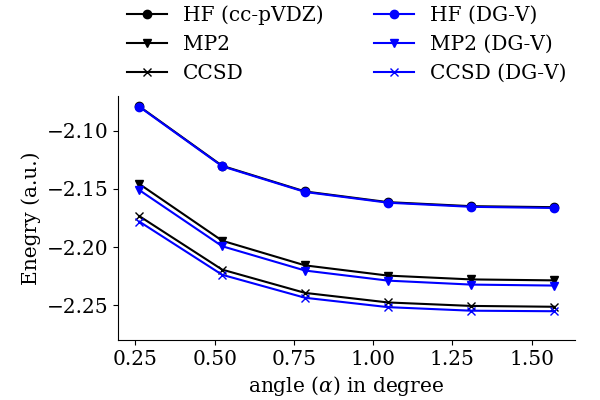}
    \caption{~}
\end{subfigure}
\begin{subfigure}{.49\textwidth}
    \centering
    \includegraphics[width=1.\textwidth]{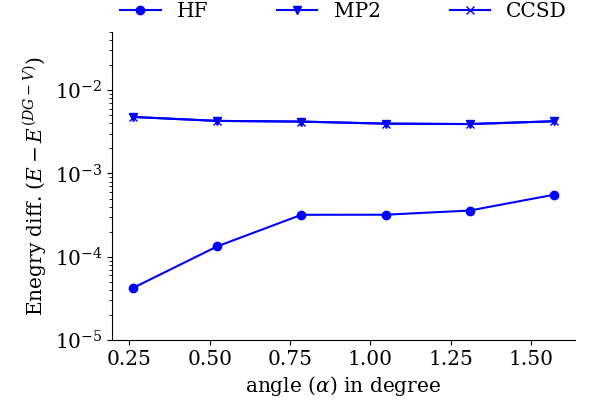}
    \caption{~}
\end{subfigure}
\begin{subfigure}{.49\textwidth}
    \centering
    \includegraphics[width=1.\textwidth]{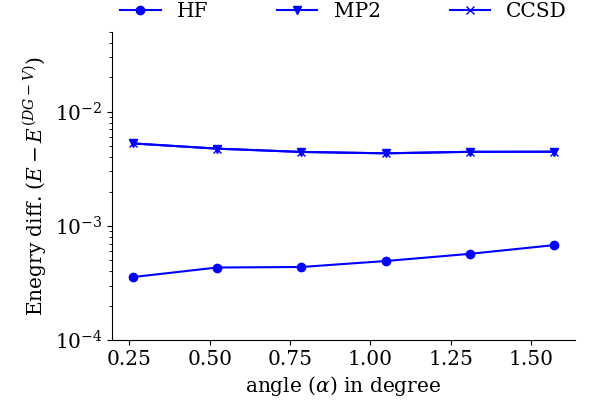}
    \caption{~}
\end{subfigure}
\caption{
Subplots (a) and (b) show the absolute energies for mean field, MP2 and CCSD energy computations. The truncations of the number of DG-V basis functions are extracted from previous mean field computations and was \emph{a priori} set to be 85\% and 90\% for aug-cc-pVDZ and cc-pVDZ, respectively. The DG-V solutions are compared to the solution provided by the periodic boundary condition module in PySCF ( depicted by the black curves).
The subplots (c) and (d) show the differences of the DG-V solutions to the respective PySCF solution, i.e. $ E_{\rm HF} - E_{\rm HF}^{\rm (DG-V)}$ on a semilogarithmic scale. }
\label{fig:correlationccpvdz}
\end{figure}

The Figures~\ref{fig:MFtruncationApp321}, \ref{fig:MFtruncationApp631} and \ref{fig:MFtruncationApp6311} show the truncation sensitivity of the Hartree--Fock energy in the DG-V basis discretization for a number of Pople basis set with added diffuse basis functions. Picking the truncation threshold for each basis set based on the calculations presented in Figures~\ref{fig:MFtruncationApp6311}, \ref{fig:MFtruncationApp6311} and \ref{fig:MFtruncationApp6311}, we can extend the results presented in Section~\ref{sec:P4toD4} on the correlated calculations for the respective basis sets. The results are presented in Figures~\ref{fig:correlation321G}, \ref{fig:correlation6311G}, and \ref{fig:correlationccpvdz}.

We here also present numerical values showing the maximal and minimal improvement by means of the DG-V procedure. 
We here highlight that the main limitation of this improvement is restricted by the choice of the active basis.

\begin{center}
\begin{table}[ht]
\centering
\begin{tabular}{r|ccccccc}%{@{}*{7}{l}}
\br
DG-V trunc. & 100\% & 95\% & 90\% & 85\% & 80\% & 70\% & 60\%  \\
\mr
    cc-pVDZ & 0.5 & 0.5 & 0.4 & -0.6 & -1.2 & -5.2 & -16.4 \\
aug-cc-pVDZ & 0.4 & 0.4 & 0.3 &  0.0 & -0.9 & -2.8 & -5.5 \\
\mr
%Basis sets & 100\% & 95\% & 90\% & 85\% & 80\% & 70\% & 60\%  \\
%\mr
3-21G & 2.4 & 2.4 & 2.4 & 2.3 & 2.3 & -0.7 & -27.2\\
3-21++G &  2.9 & 2.9 & 2.8 & 2.8 & 2.7 & 2.3 & -4.4\\
%\mr
%Basis sets & 100\% & 95\% & 90\% & 85\% & 80\% & 70\% & 60\%  \\
\mr
6-31G   & 4.1 & 4.1 & 4.1 & 4.0 & 4.0 & 0.3 & -32.2 \\
%6-31+G  & 4.1 & 4.1 & 4.1 & 4.0 & 4.0 & 0.3 & -32.2 \\
6-31++G & 4.5 & 4.5 & 4.4 & 4.4 & 4.3 & 4.3 & 0.9\\
%\mr
%Basis sets & 100\% & 95\% & 90\% & 85\% & 80\% & 70\% & 60\%  \\
\mr
6-311G   & 2.0 & 2.0 & 1.9 & 1.9 & 1.8 & 0.0 & -6.5 \\
%6-311+G  & 2.0 & 2.0 & 1.9 & 1.9 & 1.8 & 0.0 & -6.5 \\
6-311++G & 1.9 & 1.9 & 1.8 & 1.8 & 1.7 & 1.3 & -1.1\\
\br
\end{tabular}
\caption{Minimal difference in mH (min of $E_{\rm HF} - E_{\rm HF}^{\rm (DG-V)}$) along the PES between the Hartree--Fock energy in regular and DG-V discretization for different level of truncation. The truncation threshold for the number DG-V basis functions relative to the maximal number of DG-V basis functions per element is reported in percentage.}
\label{tab:MinimalDifferenceHF}
\end{table}
\end{center}

\begin{center}
\begin{table}[ht]
\centering
\begin{tabular}{r|ccccccc}%{@{}*{7}{l}}
\br
DG-V trunc. & 100\% & 95\% & 90\% & 85\% & 80\% & 70\% & 60\%  \\
\mr
    cc-pVDZ &  0.7 & 0.7 & 0.7 & 0.7 & 0.4 & 0.0 & -2.8 \\
aug-cc-pVDZ &  0.6 & 0.6 & 0.6 & 0.6 & 0.3 & 0.2 & -0.7 \\
\mr
%Basis sets & 100\% & 95\% & 90\% & 85\% & 80\% & 70\% & 60\%  \\
%\mr
3-21G & 6.7 & 6.7 & 6.7 & 6.4 & 6.4 & 2.7 & 1.1\\
3-21++G & 8.7 & 8.7 & 8.7 & 8.7 & 8.3 & 6.7 & 2.8\\
\mr
%Basis sets & 100\% & 95\% & 90\% & 85\% & 80\% & 70\% & 60\%  \\
%\mr
6-31G & 9.5 & 9.5 & 9.5 & 8.8 & 8.8 & 4.4 & 2.1 \\
%6-31+G & \\
6-31++G & 10.6 & 10.6 & 10.5 & 10.5 & 9.9 & 7.1 & 4.3\\
\mr
%Basis sets & 100\% & 95\% & 90\% & 85\% & 80\% & 70\% & 60\%  \\
%\mr
6-311G & 6.2 & 6.2 & 6.1 & 6.1 & 5.9 & 2.7 & 1.6\\
%6-311+G & \\
6-311++G & 6.1 & 6.1 &  6.1 & 6.0 & 5.9 & 5.8 & 1.8 \\
\br
\end{tabular}
\caption{Maximal difference in mH (max of $E_{\rm HF} - E_{\rm HF}^{\rm (DG-V)}$) along the PES between the Hartree--Fock energy in regular and DG-V discretization for different level of truncation. The truncation threshold for the number DG-V basis functions relative to the maximal number of DG-V basis functions per element is reported in percentage.}
\label{tab:MaximalDifferenceHF}
\end{table}
\end{center}

\FloatBarrier

\section{Hydrogen chains}

\begin{figure}[h!]
    \centering
    \begin{subfigure}{.5\textwidth}
    \centering
    \includegraphics[width=1.\textwidth]{Images/NNz_ERI_loew.png}
    \caption{~}
    \label{fig:nnz_eri_l}
    \end{subfigure}%
    \begin{subfigure}{.5\textwidth}
    \centering
    \includegraphics[width=1.\textwidth]{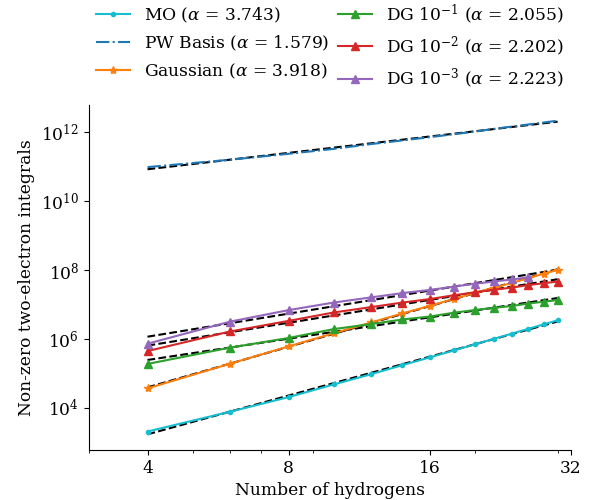}
    \caption{~}
    \label{fig:nnz_eri_nl}
\end{subfigure}
\caption{The number of nonzero two-electron integrals in different representations for SVD-truncation tolerances of $10^{-1}$, $10^{-2}$ and $10^{-3}$, plotted on a log--log scale. We fit a trendline plotted with black dots to extract the scaling as a function of system size as $N^{\alpha} + c$ for some constant $c$, and list the exponent $\alpha$ beside each representation in the legend. As predicted, for these system sizes the number of two-electron integrals lies between the primitive and active-space representations, tending closer to the $\mathcal{O}(N^2)$ scaling of the primitive representation, while requiring fewer function. In both cases we orthonormalized the Gaussian basis. In (a) we used the L\"owdin orthonormalization and in (b) SVD orthonormalization introducing an arbitrary gauge. }
\label{fig:nnz_eri_l_vs_nl}
\end{figure}

In a previous work~\cite{mcclean2020discontinuous}, the number of nonzero two-electron repulsion integrals for hydrogen chains of increasing length was compared for discretizations through Gaussian type orbitals, DG-R and planewave dual basis, aiming at an empirical cost-factor scaling analysis for quantum simulations of chemistry. To that end, the Gaussian type orbitals were orthonormalized by means of a singular values decomposition, introducing a somewhat arbitrary gauge. We here extend this study by employing the L\"owdin orthonormalization to the Gaussian type orbitals, and reporting the corresponding scaling for an active-space basis set of molecular orbitals. The considered active space of molecular orbitals (MO) consists of the $N_{\rm occ}$ occupied orbitals and the lowest $3N_{\rm occ}$ virtual orbitals. We confirm the scaling behavior of the number of nonzero two-electron repulsion integrals using an SVD orthonormalization reported in Ref.~\cite{mcclean2020discontinuous} for the considered system, see Figure~\ref{fig:nnz_eri_l_vs_nl}(b). We furthermore observe that the scaling of the number of nonzero two-electron repulsion integrals for the molecular orbitals (MO) is close to $\mathcal{O}(N^4)$. Hence, although an active space introduces a beneficial offset, extrapolation of the curves in Figure~\ref{fig:nnz_eri_l_vs_nl} show that the use of molecular orbitals becomes more expensive than DG-V beyond 74 atoms. Comparing with the results obtained using the L\"owdin orthonormalization Gaussians, see Figure~\ref{fig:nnz_eri_l_vs_nl} (a), we observe that the scaling of the number of nonzero two-electron repulsion integrals in a Gaussian atomic orbital discretization is significantly reduced. Nonetheless, we observe a crossing with a DG-V curve between 12 and 14 atoms.

%

% \begin{figure}
%     \centering
%     \includegraphics[width = 0.6\textwidth]{Images/NNz_ERI_nd.png}
%     \caption{Similar results as in the previous paper, using gto.Mole object and neglect diagonal elements in eri tensor. The number of nonzero two-electron integrals in different representations for SVD-truncation tolerances of $10^{-1}$, $10^{-2}$ and $10^{-3}$, plotted on a log–log scale. We fit a trendline plotted with black dots to extract the scaling as a function of system size as $N^{\alpha} + c$ for some constant $c$, and list the exponent $\alpha$ beside each representation in the legend. As predicted, for these system sizes the number of two-electron integrals lies between the primitive and active-space representations, tending closer to the $\mathcal{O}(N^2)$ scaling of the primitive representation,
%     requiring fewer functions}
%     \label{fig:NNZ-ERI-H-chain_coarse}
% \end{figure}

% \begin{figure}[h!]
%     \centering
%     \begin{subfigure}{.5\textwidth}
%     \centering
%     \includegraphics[width=1.\textwidth]{Images/Number_of_VdG_fcts_loew.png}
%     \caption{~}
%     \label{fig:nnz_eri_l}
%     \end{subfigure}%
%     \begin{subfigure}{.5\textwidth}
%     \centering
%     \includegraphics[width=1.\textwidth]{Images/Number_of_VdG_fcts_n_loew.png}
%     \caption{~}
%     \label{fig:nnz_eri_nl}
% \end{subfigure}
% \caption{(a)  (b) }
% \label{fig:nnz_eri_l_vs_nl}
% \end{figure}

\begin{figure}[h!]
    \centering
    \begin{subfigure}{.5\textwidth}
    \centering
    \includegraphics[width=1.\textwidth]{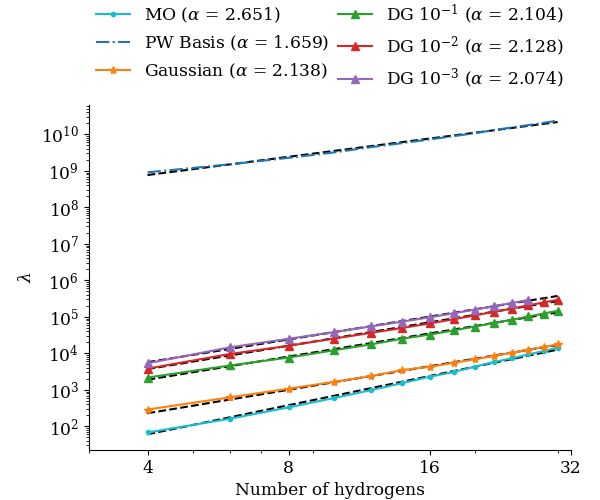}
    \caption{~}
    \label{fig:la_l}
    \end{subfigure}%
    \begin{subfigure}{.5\textwidth}
    \centering
    \includegraphics[width=1.\textwidth]{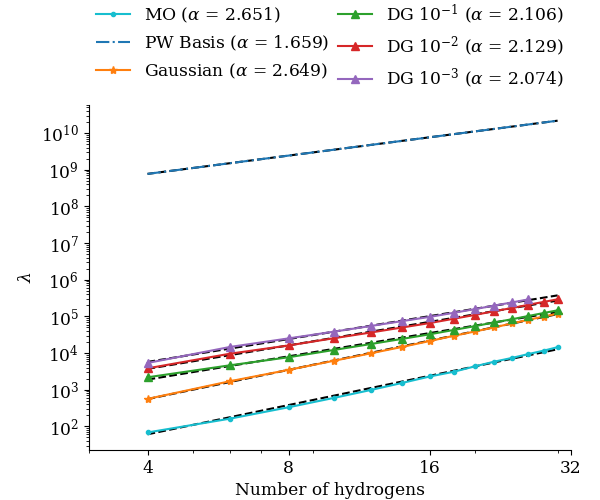}
    \caption{~}
    \label{fig:la_nl}
\end{subfigure}
\caption{$\lambda$ value for Gaussian and DG-V basis, which is related to the cost of quantum algorithms such as linear combination of unitaries (LCU). The value of $\lambda$ is plotted as a function of system size for different representations. We fit a trendline plotted with black dots from the second point onward to extract the scaling as a function of system size as $N\alpha + c$ for some constant $c$, and list the exponent $\alpha$ beside each representation in the figure legend. In (a) we used the L\"owdin orthonormalization and in (b) SVD orthonormalization introducing an arbitrary gauge. }
\label{fig:la_l_vs_nl}
\end{figure}

For certain types of quantum algorithms such as those based on the linear combination of unitaries approach~\cite{BerryChildsCleveEtAl2015,ChildsKothariSomma2017}, the major part of the cost does not come directly from the bare number of nonzero two-electron repulsion integrals but from the $\lambda$-value, i.e., the $\ell_1$-norm of the two-body interaction tensor $\lambda = \sum_{p,q,r,s}|v_{p,q,r,s}|$.
We find that the L\"owdin orthonormalization of the Gaussian type atomic orbital basis has a more significant effect to the scaling of the $\lambda$-value than on the number of nonzero two-electron repulsion integrals, see Figure~\ref{fig:la_l_vs_nl}(a). In particular, the $\lambda$ values of DG-V and active basis are almost parallel. Note however, that around 30 atoms, the $\lambda$-value for the molecular orbital basis (using an active space) crosses the L\"owdin orthonormalized Gaussian basis set, i.e., in terms of the $\lambda$-value influenced cost factors in quantum simulations, the use of L\"owdin orthonormalized Gaussians becomes more efficient compared to the use of molecular orbitals despite the use of a basis reduction through a active space consideration.

%Although the number of nonzero two-electron repulsion integrals and the $\lambda$-value appear to be directly related, the computational results provided here show that the connection is more profound than maybe expected. 
Although the improvement in the $\lambda$-value scaling through the DG formalism is not as pronounced as it is for the number of nonzero two-electron repulsion integrals, the DG procedure gives rise to a two-body interaction tensor with a block-diagonal sparsity structure. We expect that this can still significantly reduce the cost of implementing quantum algorithms, such as those based on fermionic swap networks~\cite{KivlichanMcCleanWiebeEtAl2018,mcclean2020discontinuous}.

% \begin{figure}[h!]
%     \centering
%     \includegraphics[width = 0.6\textwidth]{Images/lambda_loew.png}
%     \caption{$\lambda$ value for Gaussian and DG-V basis. A core quantity in determining the cost in quantum algorithms, $\lambda$, is plotted as a
%     function of system size for different representations. We fit a trendline plotted with black dots from the second point onward to extract the scaling as a function of system size as $N\alpha + c$ for some constant $c$, and list the exponent $\alpha$ beside each representation in the figure legend. }
%     \label{fig:lambda-H-chain}
% \end{figure}

%% file: main.bbl
\begin{thebibliography}{10}

\bibitem{Arnold1982}
D.~N. Arnold.
\newblock An interior penalty finite element method with discontinuous
  elements.
\newblock {\em SIAM J. Numer. Anal.}, 19:742 -- 760, 1982.

\bibitem{aurenhammer2013voronoi}
F.~Aurenhammer, R.~Klein, and D.-T. Lee.
\newblock {\em Voronoi diagrams and Delaunay triangulations}.
\newblock World Scientific Publishing Company, 2013.

\bibitem{babbush2018low}
R.~Babbush, N.~Wiebe, J.~McClean, J.~McClain, H.~Neven, and G.~K.-L. Chan.
\newblock Low-depth quantum simulation of materials.
\newblock {\em Phys. Rev. X}, 8(1):011044, 2018.

\bibitem{BerryChildsCleveEtAl2015}
D.~W. Berry, A.~M. Childs, R.~Cleve, R.~Kothari, and R.~Somma.
\newblock Simulating {Hamiltonian} dynamics with a truncated {Taylor} series.
\newblock {\em Phys. Rev. Lett.}, 114:090502, 2015.

\bibitem{ChildsKothariSomma2017}
A.~M. Childs, R.~Kothari, and R.~D. Somma.
\newblock Quantum algorithm for systems of linear equations with exponentially
  improved dependence on precision.
\newblock {\em SIAM J. Comput.}, 46:1920--1950, 2017.

\bibitem{ditchfield1971self}
R.~Ditchfield, W.~J. Hehre, and J.~A. Pople.
\newblock Self-consistent molecular-orbital methods. ix. an extended
  gaussian-type basis for molecular-orbital studies of organic molecules.
\newblock {\em The Journal of Chemical Physics}, 54(2):724--728, 1971.

\bibitem{dunning1989gaussian}
T.~H. Dunning~Jr.
\newblock Gaussian basis sets for use in correlated molecular calculations. i.
  the atoms boron through neon and hydrogen.
\newblock {\em The Journal of chemical physics}, 90(2):1007--1023, 1989.

\bibitem{FraserFoulkesRajagopalEtAl1996}
L.~M. Fraser, W.~M.~C. Foulkes, G.~Rajagopal, R.~Needs, S.~Kenny, and
  A.~Williamson.
\newblock {Finite-size effects and Coulomb interactions in quantum Monte Carlo
  calculations for homogeneous systems with periodic boundary conditions}.
\newblock {\em Phys. Rev. B}, 53(4):1814--1832, 1996.

\bibitem{GoedeckerTeterHutter1996}
S.~Goedecker, M.~Teter, and J.~Hutter.
\newblock Separable dual-space gaussian pseudopotentials.
\newblock {\em Phys. Rev. B}, 54:1703, 1996.

\bibitem{helgaker2014molecular}
T.~Helgaker, P.~Jorgensen, and J.~Olsen.
\newblock {\em Molecular electronic-structure theory}.
\newblock John Wiley \& Sons, 2014.

\bibitem{HuLinYang2015a}
W.~Hu, L.~Lin, and C.~Yang.
\newblock {DGDFT}: A massively parallel method for large scale density
  functional theory calculations.
\newblock {\em J. Chem. Phys.}, 143:124110, 2015.

\bibitem{jankowski1980applicability}
K.~Jankowski and J.~Paldus.
\newblock Applicability of coupled-pair theories to quasidegenerate electronic
  states: A model study.
\newblock {\em International Journal of Quantum Chemistry}, 18(5):1243--1269,
  1980.

\bibitem{KivlichanMcCleanWiebeEtAl2018}
I.~Kivlichan, J.~McClean, N.~Wiebe, C.~Gidney, A.~Aspuru-Guzik, G.-L. Chan, and
  R.~Babbush.
\newblock {Quantum Simulation of Electronic Structure with Linear Depth and
  Connectivity}.
\newblock {\em Phys. Rev. Lett.}, 120(11):110501, 2018.

\bibitem{KniziaChan2012}
G.~Knizia and G.~Chan.
\newblock Density matrix embedding: A simple alternative to dynamical
  mean-field theory.
\newblock {\em Phys. Rev. Lett.}, 109:186404, 2012.

\bibitem{KresseFurthmuller1996}
G.~Kresse and J.~Furthm{\"u}ller.
\newblock {Efficient iterative schemes for ab initio total-energy calculations
  using a plane-wave basis set}.
\newblock {\em Phys. Rev. B}, 54:11169--11186, 1996.

\bibitem{lehtola2019review}
S.~Lehtola.
\newblock A review on non-relativistic, fully numerical electronic structure
  calculations on atoms and diatomic molecules.
\newblock {\em International Journal of Quantum Chemistry}, 119(19):e25968,
  2019.

\bibitem{lin2012adaptive}
L.~Lin, J.~Lu, L.~Ying, and W.~E.
\newblock {Adaptive local basis set for Kohn-Sham density functional theory in
  a discontinuous Galerkin framework I: Total energy calculation}.
\newblock {\em J. Comput. Phys.}, 231:2140--2154, 2012.

\bibitem{Loewdin1950}
P.-O. L{\"o}wdin.
\newblock On the non-orthogonality problem connected with the use of atomic
  wave functions in the theory of molecules and crystals.
\newblock {\em J. Chem. Phys.}, 18:365--375, 1950.

\bibitem{MakovPayne1995}
G.~Makov and M.~C. Payne.
\newblock {Periodic boundary conditions in ab initio calculations}.
\newblock {\em Phys. Rev. B}, 51:4014, 1995.

\bibitem{mcclean2020discontinuous}
J.~McClean, F.~Faulstich, Q.~Zhu, B.~O'Gorman, Y.~Qiu, S.~R. White, R.~Babbush,
  and L.~Lin.
\newblock Discontinuous galerkin discretization for quantum simulation of
  chemistry.
\newblock {\em New Journal of Physics}, 2020.

\bibitem{sun2018pyscf}
Q.~Sun, T.~C. Berkelbach, N.~S. Blunt, G.~H. Booth, S.~Guo, Z.~Li, J.~Liu,
  J.~D. McClain, E.~R. Sayfutyarova, S.~Sharma, et~al.
\newblock Pyscf: the python-based simulations of chemistry framework.
\newblock {\em Wiley Interdisciplinary Reviews: Computational Molecular
  Science}, 8(1):e1340, 2018.

\bibitem{SunChan2016}
Q.~Sun and G.~K.-L. Chan.
\newblock Quantum embedding theories.
\newblock {\em Acc. Chem. Res.}, 49:2705--2712, 2016.

\end{thebibliography}
